\documentclass[pra,aps,twocolumn,superscriptaddress,showpacs]{revtex4-1}
\usepackage{graphicx,graphics,epsfig}
\usepackage{dcolumn}
\usepackage{bm}
\usepackage{amsmath}
\usepackage{verbatim}
\usepackage{color}
\usepackage[colorlinks=false]{hyperref} 
\usepackage{subfigure}
\usepackage{times,natbib}
\usepackage{amsmath,amsfonts,amssymb,graphics,graphicx,epsfig,color,times,natbib}
\usepackage{tikz}
\usepackage{pgfplots}
\usepackage{verbatim}

\date{\today}

\begin{document}
\title{Factorized Three-body S-Matrix Restrained by Yang-Baxter Equation\\ and Quantum Entanglements}

\author{Li-Wei Yu}
\email{NKyulw@gmail.com}
 \affiliation{Theoretical Physics Division, Chern Institute of Mathematics, Nankai University,
 Tianjin 300071, China}

\author{Qing Zhao}
\email{qzhaoyuping@bit.edu.cn}
 \affiliation{Physics College, Beijing Institute of Technology, Beijing 100081, China}

\author{Mo-Lin Ge}
\email{geml@nankai.edu.cn}
 \affiliation{Theoretical Physics Division, Chern Institute of Mathematics, Nankai University,
 Tianjin 300071, China}

\date{\today}


\begin{abstract}
This paper investigates the physical effects of Yang-Baxter equation (YBE) to quantum entanglements through the 3-body S-matrix in entangling parameter space. The explicit form of 3-body S-matrix $\breve{R}_{123}(\theta,\varphi)$ based on the 2-body S-matrices is given due to the factorization condition of YBE. The corresponding chain Hamiltonian has been obtained and diagonalized, also the Berry phase for 3-body system is given. It turns out that by choosing different spectral parameters the $\breve{R}(\theta,\varphi)$-matrix gives GHZ and W state respectively. The extended 1-D Kitaev toy model has been derived.  Examples of the role of the model in entanglement transfer are discussed.
\end{abstract}


\maketitle

\section{Introduction}
\label{intro}
Quantum entanglement plays an important role in quantum information due to its applications in quantum commutation and information processing\cite{bennett1993teleporting}, superdense coding\cite{bennett1992communication}, quantum key distribution\cite{ekert1991quantum} and computation\cite{nielsen2001quantum} and so on. There have been different approaches in describing quantum entanglements. On the other hand, there has been long history of Yang-Baxter equation(YBE)\cite{yang1967some,yang1968matrix,baxter2007exactly,batchelor2007bethe,yang1991braid,takhtadzhan1979quantum,faddeev1981soviet,kulish1981lecture,kulish1982solutions,korepin1997quantum} in dealing with the integrable models, low dimensional quantum field theory(QFT), statistical models and and quantum groups\cite{yang1991braid,takhtadzhan1979quantum,faddeev1981soviet,kulish1981lecture,kulish1982solutions,korepin1997quantum}. As is known well that YBE is the condition that a 3-body scattering matrix with one-dimensional spectral parameter(say, momentum) can be decomposed into three 2-body scattering matrices. Typically, YBE takes the following form:
\begin{eqnarray}\label{YBEMultiply}
&&\breve{R}_{i,i+1}(x) \breve{R}_{i+1,i+2}(xy) \breve{R}_{i,i+1}(y)\nonumber\\
&=&\breve{R}_{i+1,i+2}(y) \breve{R}_{i,i+1}(xy) \breve{R}_{i+1,i+2}(x)
\end{eqnarray}
where $\breve{R}_{i,i+1}(x)=I\otimes\cdots \underset{i,i+1}{\breve{R}(x)} \cdots I$ and $x$, $y$ stand for spectral parameters. For the familiar spin chain models usually $x=e^{iu}$, then Eq.(\ref{YBEMultiply}) becomes:
\begin{eqnarray}\label{YBEGailian}
&&\breve{R}_{i,i+1}(u) \breve{R}_{i+1,i+2}(u+v) \breve{R}_{i,i+1}(v)\nonumber\\
&=&\breve{R}_{i+1,i+2}(v) \breve{R}_{i,i+1}(u+v) \breve{R}_{i+1,i+2}(u)
\end{eqnarray}
which means that the scattering obeys the Galilean additivity for velocities $u$ and $v$. The spectral parameter independent asymptotic form of $\breve{R}(x)$ denoted by $B$ obeys braid relation:
\begin{gather}
  B_i B_{i+1}B_i=B_{i+1}B_iB_{i+1} \qquad (B_i\equiv B_{i,i+1})\label{braid relation}\\
  B_i=I\otimes\cdots \underset{i\,i+1}B \cdots I
\end{gather}
It is interesting to note that the new progress has been made to find a new type of solutions of Eq.(\ref{braid relation}) different from those stated at the beginning of this section that is traditionally familiar and called Type-I. For simplicity we call the ``traditional'' solutions of Eq.(\ref{braid relation}) Type-I, whereas the new type of solutions related to quantum entanglement are called Type-II.

The Type-II of braiding matrices satisfying Eq.(\ref{braid relation}) was first proposed in Ref.\cite{dye2003unitary}, then extended to the corresponding matrices obeying YBE(Eq.\ref{YBEMultiply}) which no longer takes the form(Eq.\ref{YBEGailian}). For the self-contain we briefly summarize the main results for the Type-II solutions of Eq.(\ref{YBEMultiply}) and (\ref{braid relation}). The key observation is that the matrix $B_{i,i+1}$ transforming the natural basis $|\psi_0\rangle$ to the Bell states $|\psi_{B}\rangle$ obeys the braid relation(Eq.\ref{braid relation})\cite{kauffman2004braiding}.
\begin{equation}
  |\psi_{B}\rangle=B_{i,i+1}\left(
  \begin{array}{c}
  |00\rangle \\|01\rangle \\|10\rangle \\|11\rangle
  \end{array}
  \right)\equiv B_{i,i+1}|\psi_0\rangle
\end{equation}

\begin{equation}\label{type II}
   B_{i,i+1} = \tfrac{1}{\sqrt{2}}\left[
   \begin{array}{cccc}
   1 & 0 & \ 0 & \ e^{i\varphi} \\
   0 & 1 & \ 1 & \ 0 \\
   0 & -1 & \ 1 & \ 0 \\
   -e^{-i\varphi} & 0 & \ 0 & \ 1
   \end{array} \right] = \tfrac{1}{\sqrt{2}}\left(I+M\right)
\end{equation}
where $M^2=-I$, and
 \begin{equation}\label{braiding matrix}
  B_{i,i+1}\left(
  \begin{array}{c}
  |00\rangle \\|01\rangle \\|10\rangle \\|11\rangle
  \end{array}
  \right)=\tfrac{1}{\sqrt{2}}\left(
  \begin{array}{c}
  |00\rangle-e^{-i\varphi}|11\rangle \\|01\rangle-|10\rangle \\|10\rangle+|01\rangle \\e^{i\varphi}|00\rangle+|11\rangle
  \end{array}
  \right)
\end{equation}
Further Eq.(\ref{braiding matrix}) can be parametrized to satisfy YBE(Eq.\ref{YBEMultiply}) in terms of the new spectral parameter $\theta$:
\begin{equation}
\breve{R}_{i,i+1}=\tfrac{1}{\sqrt{1+x^2}}(B_{i,i+1}+xB_{i,i+1}^{-1})
\end{equation}
by introducing new variable of angle $\theta$ as $\cos\theta=\tfrac{1+x}{\sqrt{2(1+x^2)}}$ and $\sin\theta=\tfrac{1-x}{\sqrt{2(1+x^2)}}$, we have:
\begin{equation}\label{Type-II}
   \breve{R}_{i,i+1}(\theta,\varphi) = \left[
   \begin{array}{cccc}
   \cos\theta & 0 & \ 0 & \ \sin\theta e^{i\varphi} \\
   0 & \cos\theta & \ \sin\theta & \ 0 \\
   0 & -\sin\theta & \ \cos\theta & \ 0 \\
   -\sin\theta e^{-i\varphi} & 0 & \ 0 & \ \cos\theta
   \end{array} \right]
\end{equation}
In Ref.\cite{chen2007braiding,zhang2005universal,hu2008optical}, 2-body S-matrix $\breve{R}_{i,i+1}(\theta,\varphi)$ has been introduced from the Yang-Baxterization approach for discussing the entanglement of 2-qubit pure states. By acting the unitary matrix $\breve{R}_{i,i+1}(\theta,\varphi)$  on the direct product state $|kl\rangle=|k\rangle\otimes|l\rangle$, we get the four states:
\begin{equation}
  \breve{R}_{i,i+1}\left(
  \begin{array}{c}
  |00\rangle \\|01\rangle \\|10\rangle \\|11\rangle
  \end{array}
  \right)=\left(
  \begin{array}{c}
  \cos\theta|00\rangle-\sin\theta e^{-i\varphi}|11\rangle \\\cos\theta|01\rangle-\sin\theta|10\rangle \\\sin\theta|01\rangle+\cos\theta|10\rangle \\\sin\theta e^{i\varphi}|00\rangle+\cos\theta|11\rangle
  \end{array}
  \right)
\end{equation}
The physical meaning of $\theta$ is related to the entanglement degree, from the right hand side of the above equation, the four states process the same degree of entanglement with $|\sin2\theta|$: when $\theta=\frac{\pi}{4}$, $\breve{R}_{i,i+1}(\theta,\varphi)$ turns to the braiding operator and four states reduce to Bell basis.

For the self-contain we briefly summarize the known results for the connection between YBE and quantum entanglements.

\smallskip

(A) Setting $\varphi=0$ in Eq.(\ref{type II}), $M$ becomes $M=i\gamma_1$  where $\gamma_1$ is Dirac matrix. Because $\gamma_{1}=-C$ where $C$ is the charge conjugate operator in Majorana representation. Thus Eq.(\ref{Type-II}) leads to:
\begin{equation}
S(\theta)=\breve{R}(\theta)=e^{-i\theta C}
\end{equation}
whose eigenstates are $S(\theta)|\xi_{\pm}\rangle=e^{\pm i\theta}|\xi_{\pm}\rangle$ and $S(\theta)|\eta_{\pm}\rangle=e^{\pm i\theta}|\eta_{\pm}\rangle$ with $|\xi_{\pm}\rangle=\frac{1}{\sqrt{2}}(|\uparrow\uparrow\rangle\pm i|\downarrow\downarrow\rangle)$ and $|\eta_{\pm}\rangle=\frac{1}{\sqrt{2}}(|\uparrow\downarrow\rangle\pm i|\downarrow\uparrow\rangle)$.

\smallskip

(B) To satisfy the YBE parameterized in terms of $\theta_1$, $\theta_2$ and $\theta_3$
 \begin{eqnarray}\label{YBEcondition}
    &&\breve{R}_{12}(\theta_1,\varphi) \breve{R}_{23}(\theta_2,\varphi) \breve{R}_{12}(\theta_3,\varphi)\nonumber\\
    &=&\breve{R}_{23}(\theta_3,\varphi) \breve{R}_{12}(\theta_2,\varphi) \breve{R}_{23}(\theta_1,\varphi)
 \end{eqnarray}
 the condition holds

\begin{equation}\label{lorentzadd}
   \tan\theta_2=\frac{\tan\theta_1+\tan\theta_3}{1+\tan\theta_1 \tan\theta_3}
\end{equation}
If $u=\tan\theta$, Eq.(\ref{lorentzadd}) reads the Lorentz additivity for $c=1$ rather than Galilean.

(C) The Hamiltonian associated with 2-body entanglement takes the form for $\varphi=\varphi(t)$:
\begin{eqnarray}
 \hat{H}_{i,i+1}&=&-\hbar \dot{\varphi}\sin\theta [\tfrac{\sin\theta}{2}\left(S_i^3+S_{i+1}^3\right)\nonumber\\
 &&+\cos\theta\left(e^{i\varphi}S_i^+S_{i+1}^++e^{-i\varphi}S_i^-S_{i+1}^-\right)]
\end{eqnarray}
i.e. the 1D Kitaev model without hopping term\cite{kitaev2001unpaired}. The corresponding eigenvalues and eigenstates for $\varphi=\Omega t$ are:
\begin{subequations}
\begin{align*}
&E_{\pm}=\mp \hbar \Omega \sin\theta\\
&|\psi_+\rangle(\theta,\varphi) =\cos(\tfrac{\pi}{4}-\tfrac{\theta}{2})\left|\uparrow\uparrow\right\rangle +\sin(\tfrac{\pi}{4}-\tfrac{\theta}{2})e^{-i\varphi}\left|\downarrow\downarrow\right\rangle\\
&|\psi_-\rangle(\theta,\varphi) =-\sin(\tfrac{\pi}{4}-\tfrac{\theta}{2})e^{i\varphi}\left|\uparrow\uparrow\right\rangle +\cos(\tfrac{\pi}{4}-\tfrac{\theta}{2})\left|\downarrow\downarrow\right\rangle
\end{align*}
\end{subequations}
The corresponding Berry phase is\cite{chen2007braiding,zhang2005universal,hu2008optical}
\begin{equation}
\gamma_{\pm}=\pm\pi(1-\sin\theta)
\end{equation}

\medskip

  There are different approaches in describing 3-qubit entanglement for pure states\cite{dur2000three,PhysRevLett.92.087902,chen2004gisin,han2004compatible,zhao2013identification}. Inspired by the relationship between 2-body S-matrix and 2-qubit entanglement, investigating how 3-body S-matrix is related to 3-qubit pure state entanglement turns interesting. If a 3-body S-matrix related to 3-qubit entanglement can be decomposed into three 2-body S-matrices as constrained by YBE, we then express the 3-qubit entanglement in terms of three 2-qubit entanglements explicitly. Since from the view-point of S-matrix theory, it is acceptable to regard 2-body scattering as basic ones in low energy phenomena. Suppose a 3-body S-matrix can be expressed by
 \begin{eqnarray}
 &&\breve{R}_{123}(\theta_1,\theta_2,\theta_3,\varphi)\\
 &=&\breve{R}_{12}(\theta_1,\varphi) \breve{R}_{23}(\theta_2,\varphi) \breve{R}_{12}(\theta_3,\varphi)\nonumber\\
&=&\breve{R}_{23}(\theta_3,\varphi) \breve{R}_{12}(\theta_2,\varphi) \breve{R}_{23}(\theta_1,\varphi)\nonumber
 \end{eqnarray}
The relation for $\theta_1$, $\theta_2$ and $\theta_3$ is then
\begin{equation}\label{Anglerelation}
  \tan\theta_1 \tan\theta_2 \tan\theta_3 +\tan\theta_2-\tan\theta_1-\tan\theta_3=0
\end{equation}
Namely, $\theta_2$ can be replaced by $\theta_1$ and $\theta_3$ by applying Eq. (\ref{Anglerelation}) in $\breve{R}_{123}(\theta_1,\theta_2,\theta_3,\varphi)$, then $\breve{R}_{123}$  depends on $\theta_1$, $\theta_3$ and $\varphi$ only.

The calculation shows that $\breve{R}_{123}$ can be expressed in the form(see Appendix A):
\begin{equation}\label{3Smatrix}
  \breve{R}_{123}(\eta,\beta,\varphi)=e^{i\eta \left(\vec{n}\cdot\vec{\Sigma}\right)}
\end{equation}
where

\begin{eqnarray*}
  \cos\eta &=& \cos\theta_2 \cos\left(\theta_1+\theta_3\right)\\
  \sin\eta &=& \sin\theta_2\sqrt{1+\cos^2(\theta_1-\theta_3)}\\
  \vec{n} &=& \left(
  \begin{array}{ccc}
  \tfrac{1}{\sqrt{2}}\cos\beta,& \tfrac{1}{\sqrt{2}}\cos\beta,& \sin\beta
  \end{array}\right)\\
  \vec{\Sigma} &=& \left(
  \begin{array}{ccc}
  \Sigma_1 ,& \Sigma_2 ,& \Sigma_3
  \end{array}\right)\nonumber\\
  &=& \left(
  \begin{array}{ccc}
  \sigma_2\otimes\sigma_1\otimes I,& I\otimes\sigma_2\otimes\sigma_1,& \sigma_2\otimes\sigma_3\otimes\sigma_1
  \end{array}\right)\\
  \cos\beta &=& \tfrac{\sqrt{2}\cos\left(\theta_1-\theta_3\right)}{\sqrt{1+\cos^2\left(\theta_1-\theta_3\right)}}\\
  \sin\beta &=& \tfrac{-\sin\left(\theta_1-\theta_3\right)}{\sqrt{1+\cos^2\left(\theta_1-\theta_3\right)}}\\
 \end{eqnarray*}
and $\sigma_1=\left(\begin{array}{cc}
  0 & e^{i\varphi} \\
  e^{-i\varphi} & 0
  \end{array}\right)
$, $\sigma_2=\left(\begin{array}{cc}
  0 & -ie^{i\varphi} \\
  ie^{-i\varphi} & 0
  \end{array}\right)$, $\sigma_3=\left(\begin{array}{cc}
  1 & 0 \\
  0 & -1
  \end{array}\right)$. It is easy to check that $\Sigma_1$, $\Sigma_2$ and $\Sigma_3$ satisfy $[\Sigma_i, \Sigma_j]=i\epsilon_{ijk}\Sigma_k(i,j,k=1,2,3)$ with $\Sigma^2=3/4$.

\medskip

For $\breve{R}_{12}=e^{i\theta M}$ with $M=\sigma_2\otimes\sigma_1$ , $\theta$ defines the entangled degree of 2-qubit system. In analogy to $\theta$, here $\eta\left(\theta_1+\theta_3, \theta_1-\theta_3\right)$ and $\beta\left(\theta_1-\theta_3\right)$ are used to define the entangled degree of certain type 3-qubit pure states. In dealing with 2-body S-matrix $\breve{R}_{12}$ and quantum entanglement, the corresponding Hamiltonian was obtained and diagonalized, and the Berry phase had been also calculated. In this paper, we extend the treatments for 2-body case to 3-body S-matrix. Moreover, the Hamiltonian of chain model induced from 3-body S-matrix can be derived in comparison to 1D Kitaev toy model\cite{kitaev2001unpaired} which generates unpaired Majorana fermions at the end of the chain. Our aim in this work is first to give explicit description of 3-qubit entanglement in terms of the known 2-qubit ones due to YBE. Further, in comparison with the polytope model of entanglement in Refs.\cite{han2004compatible,walter2013entanglement}, the constrain of YBE looks a section of the polytope in Ref.\cite{han2004compatible}. Then the corresponding Berry phase is given. The Hamiltonian for 3-body S-matrix is calculated which is the 1D Kitaev model with next nearest neighbouring hopping term. Finally we present the role of our chain model in the entanglement transfer.

\medskip

The paper is organized as follows: In Sec.\ref{Sec:2}, the unitary operator $\breve{R}_{123}(\eta,\beta,\varphi)$ is obtained, that acts on 3-qubit natural basis and generates states related to GHZ states and W states, then we give the constrain of YBE in the entanglement polytope; In Sec.\ref{Sec:3}, the Hamiltonian constructed in terms of the unitary
 matrix $\breve{R}_{123}(\eta,\beta,\varphi)$ is obtained, where $\varphi$ is time-dependent and $\eta$, $\beta$ time-independent. Based on
 the eigenstates of Hamiltonian, the Berry phase in the entanglement space is investigated; In Sec.\ref{sec:4},  we briefly discuss the relationship between 2-body S-matrix and 1D Kitaev toy model\cite{kitaev2001unpaired}, then we discuss the chain model for 3-body S-Matrix; In Sec.\ref{Sec:5}, a particular example of the chain model is discussed in the respect of entanglement transfer. The conclusion and discussion will be made in the latest section.

\section{3-body S-matrix transformation and discussion}\label{Sec:2}

Through the discussion in Refs.\cite{chen2007braiding,zhang2005universal} that the entangled degree is directly related to parameter $\theta$ for 2-body S-matrix $\breve{R}_{12}$, because $\sin2\theta$ can be defined to describe the entangled degree of 2-qubit state. Similarly, for 3-body S-matrix(Eq.\ref{3Smatrix}) acting on the direct product states, i.e. the natural basis $|klm\rangle=|k\rangle\otimes|l\rangle\otimes|m\rangle$, we have the following form
\scriptsize
\begin{eqnarray}\label{3bodytransformation}
  &&\breve{R}_{i,i+1,i+2}\left(\eta, \beta, \varphi\right)\left[\begin{array}{c}
  |000\rangle \\|001\rangle \\|010\rangle \\|011\rangle \\|100\rangle \\|101\rangle \\|110\rangle \\|111\rangle
  \end{array}\right]=\\
  &&\left[
  \begin{array}{c}
  \cos\eta|000\rangle-\frac{\cos\beta \sin\eta }{\sqrt{2}e^{i2\varphi}}|011\rangle-\frac{\sin\beta \sin\eta}{e^{i2\varphi}} |101\rangle-\frac{\cos\beta \sin\eta }{\sqrt{2}e^{i2\varphi}}|110\rangle\\
  \cos\eta|001\rangle-\frac{\cos\beta \sin\eta}{\sqrt{2}}|010\rangle-\sin\beta \sin\eta |100\rangle-\frac{\cos\beta \sin\eta }{\sqrt{2}e^{i2\varphi}}|111\rangle\\
  \frac{\cos\beta \sin\eta}{\sqrt{2}}|001\rangle+\cos\eta|010\rangle-\frac{\cos\beta \sin\eta}{\sqrt{2}} |100\rangle+\frac{\sin\beta \sin\eta }{e^{i2\varphi}}|111\rangle\\
  \frac{\cos\beta \sin\eta }{\sqrt{2}e^{-i2\varphi}}|000\rangle+\cos\eta|011\rangle-\frac{\cos\beta \sin\eta}{\sqrt{2}} |101\rangle+\sin\beta \sin\eta |110\rangle\\
  \sin\beta \sin\eta|001\rangle+\frac{\cos\beta \sin\eta}{\sqrt{2}}|010\rangle+\cos\eta |100\rangle-\frac{\cos\beta \sin\eta }{\sqrt{2}e^{i2\varphi}}|111\rangle\\
  \frac{\sin\beta \sin\eta }{e^{-i2\varphi}}|000\rangle+\frac{\cos\beta \sin\eta}{\sqrt{2}}|011\rangle+ \cos\eta |101\rangle-\frac{\cos\beta \sin\eta}{\sqrt{2}}|110\rangle\\
  \frac{\cos\beta \sin\eta }{\sqrt{2}e^{-i2\varphi}}|000\rangle-\sin\beta \sin\eta|011\rangle+\frac{\cos\beta \sin\eta}{\sqrt{2}} |101\rangle+\cos\eta|110\rangle\\
  \frac{\cos\beta \sin\eta}{\sqrt{2}e^{-i2\varphi}}|001\rangle-\frac{\sin\beta \sin\eta}{e^{-i2\varphi}}|010\rangle+\frac{\cos\beta \sin\eta}{\sqrt{2}e^{-i2\varphi}} |100\rangle+\cos\eta|111\rangle\nonumber
  \end{array}\right] \quad
\end{eqnarray}
\normalsize
Now let us discuss the entangled degree of above states. In Ref.\cite{zhao2013identification}, genuine entanglement of 3-qubit pure state is identified. For a bipartite pure state $|\phi\rangle$, its concurrence is defined by $C\left(|\phi\rangle\right)=\sqrt{1-tr\rho_1^2}$, with $\rho_1=tr_2\left(|\phi\rangle\langle\phi|\right)$. And for 3-qubit pure state $|\phi\rangle=\Sigma_{i,j,k=0}^{1}a_{ijk}|i,j,k\rangle$, $\Sigma_{i,j,k=0}^{1}|a_{ijk}|^2=1$, if the 3-qubit pure state is viewed as a bipartite state under the partition of one qubit and the rest qubits, the squared concurrence can be three cases:
\begin{subequations}\label{concurrence}
 \begin{align}
 &\scriptstyle C_{1|23}^2=\left(\Sigma_{j,k=0}^{1}|a_{0jk}|^2\right)\left(\Sigma_{j,k=0}^{1}|a_{1jk}|^2\right)-\left|\Sigma_{j,k=0}^{1}a_{0jk}a_{1jk}^{\ast}\right|\\
 &\scriptstyle C_{2|13}^2=\left(\Sigma_{i,k=0}^{1}|a_{i0k}|^2\right)\left(\Sigma_{i,k=0}^{1}|a_{i1k}|^2\right)-\left|\Sigma_{i,k=0}^{1}a_{i0k}a_{i1k}^{\ast}\right|\\
 &\scriptstyle C_{3|12}^2=\left(\Sigma_{i,j=0}^{1}|a_{ij0}|^2\right)\left(\Sigma_{i,j=0}^{1}|a_{ij1}|^2\right)-\left|\Sigma_{i,j=0}^{1}a_{ij0}a_{ij1}^{\ast}\right|
 \end{align}
\end{subequations}
For any 3-qubit pure state $|\phi\rangle$, it is genuine entangled if and only if its concurrences for all bipartite partitions are not zero. $|\phi\rangle$ is biseprable if and only if one concurrence is zero, and the other two concurrences are non-zero\cite{zhao2013identification}. If all of the concurrences are zero, $|\phi\rangle$ is fully separable.
All the states in the right hand side of Eq.(\ref{3bodytransformation}) share the same concurrences:
\begin{subequations}\label{YBEConcurrence}
\begin{align}
&\scriptstyle C_{1|23}^2=\left(\cos^2\eta+\tfrac{1}{2}\cos^2\beta \sin^2\eta\right)\left(\tfrac{1}{2}\cos^2\beta \sin^2\eta+\sin^2\beta \sin^2\eta\right)\\
&\scriptstyle C_{2|13}^2=\left(\cos^2\eta+\sin^2\beta \sin^2\eta\right)\left(\cos^2\beta \sin^2\eta\right)\\
&\scriptstyle C_{3|12}^2=\left(\cos^2\eta+\tfrac{1}{2}\cos^2\beta \sin^2\eta\right)\left(\tfrac{1}{2}\cos^2\beta \sin^2\eta+\sin^2\beta \sin^2\eta\right)
\end{align}
\end{subequations}
 When $\eta=\tfrac{\pi}{3}$, $\cos\beta=-\tfrac{\sqrt{6}}{3}$, $\sin\beta=-\tfrac{\sqrt{3}}{3}$, $\varphi=0$, the three concurrences $C_{1|23}^2=C_{2|13}^2=C_{3|12}^2=\tfrac{1}{4}$ are maximal simultaneously. Then the transformed states  $\breve{R}_{i,i+1,i+2}|\Psi_0\rangle\equiv|\Psi\rangle$ are:
\begin{equation}
\scriptstyle
  \breve{R}_{i,i+1,i+2}\left[\begin{array}{c}
  |000\rangle \\|001\rangle \\|010\rangle \\|011\rangle \\|100\rangle \\|101\rangle \\|110\rangle \\|111\rangle
  \end{array}\right]
  =\frac{1}{2}\left[
  \begin{array}{c}
  |000\rangle+|011\rangle+|101\rangle+|110\rangle\\
  |001\rangle+|010\rangle+|100\rangle+|111\rangle\\
  -|001\rangle+|010\rangle+|100\rangle-|111\rangle\\
  -|000\rangle+|011\rangle+|101\rangle-|110\rangle\\
  -|001\rangle-|010\rangle+ |100\rangle+|111\rangle\\
  -|000\rangle-|011\rangle+ |101\rangle+|110\rangle\\
  -|000\rangle+|011\rangle-|101\rangle+|110\rangle\\
  -|001\rangle+|010\rangle-|100\rangle+|111\rangle
  \end{array}\right]
\end{equation}
These states are related to GHZ states under local unitary transformation:
\begin{equation}\label{GHZstates}
H\otimes H\otimes H |\Psi\rangle =\frac{1}{\sqrt{2}}\left[
  \begin{array}{c}
  |000\rangle+|111\rangle\\
  |000\rangle-|111\rangle\\
  |001\rangle-|110\rangle\\
  -|001\rangle-|110\rangle\\
  |011\rangle-|100\rangle\\
  -|011\rangle-|100\rangle\\
  -|010\rangle-|101\rangle\\
  -|010\rangle+|101\rangle
  \end{array}\right]=|GHZ\rangle
\end{equation}
where $H$ is Hadamard gate, $H=\frac{1}{\sqrt{2}}\left[\begin{array}{cc} 1 & 1\\1 & -1 \end{array}\right] $.

In comparison with 2-qubit entangled states generated by 2-body S-matrix $\breve{R}_{i,i+1}$, we find that they are very similar to each other. Four 2-qubit entangled states share the same concurrence $|\sin2\theta|$ and become Bell basis when the concurrence is maximal, and eight 3-qubit entangled states have the same concurrence defined in Eq.(\ref{YBEConcurrence}) and become the states related to GHZ states when all of the three concurrences are maximal.

As mentioned in Ref.\cite{dur2000three}, there are two types of genuine entangled states under stochastic local operations and classical
communication(SLOCC), $|GHZ\rangle=\tfrac{1}{\sqrt{2}}(|000\rangle+|111\rangle)$ and
$|W\rangle = \tfrac{1}{\sqrt{3}}(|001\rangle+|010\rangle+|100\rangle)$. Having generated states related to GHZ states, it is challenging to generate
another type of genuine entangled W state by 3-body S-matrix $\breve{R}_{123}$. By choosing $\eta=\frac{\pi}{2}$, $\cos\beta=-\frac{\sqrt{6}}{3}$,
$\sin\beta=-\frac{\sqrt{3}}{3}$ and $\varphi=0$, we get:
\begin{equation}\label{Wstates}
  \breve{R}_{i,i+1,i+2}\left[\begin{array}{c}
  |000\rangle \\|001\rangle \\|010\rangle \\|011\rangle \\|100\rangle \\|101\rangle \\|110\rangle \\|111\rangle
  \end{array}\right]
  =\frac{1}{\sqrt{3}}\left[
  \begin{array}{c}
  |011\rangle+|101\rangle+|110\rangle\\
  |010\rangle+|100\rangle+|111\rangle\\
  -|001\rangle+|100\rangle-|111\rangle\\
  -|000\rangle+|101\rangle-|110\rangle\\
  -|001\rangle-|010\rangle+|111\rangle\\
  -|000\rangle-|011\rangle+|110\rangle\\
  -|000\rangle+|011\rangle-|101\rangle\\
  -|001\rangle+|010\rangle-|100\rangle
  \end{array}\right]=|\Phi\rangle
\end{equation}
These eight states are W states that represent another type of genuine entangled state in 3-qubit pure state system, and the concurrences of them are $C_{1|23}^2=C_{2|13}^2=C_{3|12}^2=\frac{2}{9}$.

\medskip

Now let us discuss the generated 3-qubit states via the approach of entanglement polytopes \cite{han2004compatible,walter2013entanglement} that detect multiparticle entanglement from single-particle information. According to Ref.\cite{han2004compatible}, N one-party reduced density matrices $\rho_{i}$ of an N-qubit pure state are obtained and the smaller eigenvalues $\lambda(i)$ of the one-party matrices $\rho_{i}$ obey the following sufficient and necessary condition:
\begin{equation}\label{inequality}
\sum_{i\neq k,i=1}^{N}\lambda(i)\geq \lambda(k)(k=1,2,\cdots,N)
\end{equation}
For 3-qubit pure states, Eq.(\ref{inequality}) and the normalization condition form the polytope in $\lambda_{i}$(i=1,2,3) space. Taking the state generated by 3-body S-matrix $|\psi\rangle=\breve{R}_{123}\left(\eta, \beta, \varphi\right)|000\rangle$ as an example,
\begin{eqnarray*}
|\psi\rangle  &=&\cos\eta|000\rangle-\tfrac{\cos\beta\sin\eta}{\sqrt{2}e^{i2\varphi}}|011\rangle\\
&&-\tfrac{\sin\beta\sin\eta}{e^{i2\varphi}} |101\rangle-\tfrac{\cos\beta\sin\eta}{\sqrt{2}e^{i2\varphi}}|110\rangle\\
&=&A_1|000\rangle+\tfrac{A_2}{\sqrt{2}}|011\rangle+A_3|101\rangle+\tfrac{A_2}{\sqrt{2}}|110\rangle \\
&&(|A_1|^2+|A_2|^2+|A_3|^2=1)
\end{eqnarray*}
We then have three smaller eigenvalues of one-party reduced density matrices from $|\psi\rangle$:
\begin{subequations}\label{000state}
\begin{align}
\lambda(1)&=\min\{|A_1|^2+\tfrac{|A_2|^2}{2},|A_3|^2+\tfrac{|A_2|^2}{2}\}\\
\lambda(2)&=\min\{|A_1|^2+|A_3|^2,|A_2|^2\}\\
\lambda(3)&=\min\{|A_1|^2+\tfrac{|A_2|^2}{2},|A_3|^2+\tfrac{|A_2|^2}{2}\}
\end{align}
\end{subequations}
In Eq.(\ref{000state}), $\lambda(1)=\lambda(3)$, which means that $|\psi\rangle$ must be located in the cross section of the polytope in Ref.\cite{han2004compatible}. This is because of the constrain due to YBE.

\medskip

Let us express 3-qubit concurrence in terms of $\theta_1$, $\theta_2$ and $\theta_3$. The 3-body S-matrix can be decomposed into three 2-body S-matrices as constrained by YBE, each 2-body S-matrix corresponds to the entangled degree of the 2-qubit state. Now we set up the relationship between 3-qubit entanglement and 2-qubit entanglements. By acting $\breve{R}_{123}$ on $|\Psi_{0}\rangle=|0\rangle_{1}\otimes|0\rangle_{2}\otimes|0\rangle_{3}$, we have:
\begin{eqnarray}
 |\tilde{\psi}\rangle\!&=&\!\breve{R}_{123}(\theta_1,\theta_2,\theta_3,\varphi)|000\rangle\nonumber\\
 &=&\breve{R}_{12}(\theta_1,\varphi)\breve{R}_{23}(\theta_2,\varphi)\breve{R}_{12}(\theta_3,\varphi)|000\rangle\nonumber\\
 &=&\cos\theta_2\cos(\theta_1+\theta_3)|000\rangle\nonumber\\
 &&-e^{-i\varphi}\sin\theta_2\cos(\theta_1-\theta_3)|011\rangle\nonumber\\
 &&+e^{-i\varphi}\sin\theta_2\sin(\theta_1-\theta_3)|101\rangle\nonumber\\
 &&-e^{-i\varphi}\cos\theta_2\sin(\theta_1+\theta_3)|110\rangle\nonumber\\
 &=&B_1|000\rangle+B_2|011\rangle+B_3|101\rangle+B_4|110\rangle\label{OriginState}
 \end{eqnarray}
with the Yang-Baxter equation condition(see Eq.\ref{Anglerelation}) in alternative form£º
\begin{equation}\label{YBEConSin}
\frac{\sin\theta_2}{\cos\theta_2}=\frac{\sin(\theta_1+\theta_3)}{\cos(\theta_1-\theta_3)}
\end{equation}
the generated state can be recast to:
\begin{eqnarray}
 |\tilde{\psi}\rangle\negthinspace&=&\negthinspace\cos\theta_2\cos(\theta_1+\theta_3)|000\rangle\nonumber\\
 &&-e^{-i\varphi}\sin\theta_2\cos(\theta_1-\theta_3)|011\rangle\nonumber\\
 &&+e^{-i\varphi}\sin\theta_2\sin(\theta_1-\theta_3)|101\rangle\nonumber\\
 &&-e^{-i\varphi}\sin\theta_2\cos(\theta_1-\theta_3)|110\rangle\nonumber\\
 \negthinspace&=&\negthinspace B_1|000\rangle+B_2|011\rangle+B_3|101\rangle+B_2|110\rangle
 \end{eqnarray}
 According to the definition of concurrence(Eq.\ref{concurrence}) for 3-qubit states, we express the concurreences in terms of $\sin2\theta_1$, $\sin2\theta_2$ and $\sin2\theta_3$, whose absolute value represent entangled degree of 2-qubit state when $\breve{R}_{i,i+1}(\theta)$ acts on 2-qubit direct product state. Then the resultant concurrences are:
\begin{subequations}
 \begin{align}
 &C_{1|23}^2=C_{3|12}^2=\tfrac{1}{4}|\sin2\theta_2|^2\\
 &C_{2|13}^2=-\tfrac{1}{4}\left[\sin2\theta_2(\sin2\theta_1+\sin2\theta_3)-1\right]^2+\tfrac{1}{4}
 \end{align}
\end{subequations}
with the constrain
\begin{equation}\label{ConcurrenceRelation}
\sin2\theta_2=\frac{\sin2\theta_1+\sin2\theta_3}{1+\sin2\theta_1\sin2\theta_3}
\end{equation}
For $0\leq\theta_1,\theta_2,\theta_3\leq \pi/2$,  Eq.(\ref{ConcurrenceRelation}) means that the concurrence satisfy Lorentz addition rather than the Galilean. We find that $C_{1|23}^2$ and $C_{3|12}^2$ only depends on $|\sin2\theta_2|$, whereas $C_{2|13}^2$ depends on both $\sin2\theta_1$ and $\sin2\theta_3$.

Let us consider two types of genuine entangled 3-qubit states:

\smallskip

For GHZ state, the condition is $C_{1|23}^2=C_{2|13}^2=C_{3|12}^2=\frac{1}{4}$. Then we get $|\sin2\theta_2|=1$, $|\sin2\theta_1|=1$ and$|\sin2\theta_3|=0$(or $|\sin2\theta_1|=0$ and$|\sin2\theta_3|=1$). Clearly that when 3-qubit state is GHZ state under the local unitary transformation, not all the 2-qubit concurrences($|\sin2\theta_1|$, $|\sin2\theta_2|$ and $|\sin2\theta_3|$) are maximal.

For W state, the condition is $C_{1|23}^2=C_{2|13}^2=C_{3|12}^2=\frac{2}{9}$. and the 2-qubit concurrences $|\sin2\theta_2|=\frac{2\sqrt{2}}{3}$, $|\sin2\theta_1|=|\sin2\theta_3|=\frac{\sqrt{2}}{2}$. The three 2-qubit concurrences are also not maximal.

When all of the 2-qubit concurrences are maximal, i.e. $\sin2\theta_1=\sin2\theta_2=\sin2\theta_3=1$, the generated state is biseparable($C_{2|13}^2=0$): $-\frac{\sqrt{2}e^{-i\varphi}}{2}(|011\rangle+|110\rangle)$.

Thus we show that the values of 2-qubit concurrences generate the two types of 3-qubit genuine entangled states, GHZ state and W state. If 3-body S-matrix can be decomposed into three 2-body S-matrices restricted by YBE, the generated 3-qubit state is biseparable when the 2-qubit concurrences of the three 2-body S-matrices are maximal.

\section{Hamiltonian for 3-body S-matrix and Berry phase in entanglement space}
\label{Sec:3}

As has been shown in Eq.(\ref{3bodytransformation}), there are three parameters $\eta$, $\beta$ and $\varphi$ in the 3-body S-matrix. If only $\varphi$ is time-dependent, the Hamiltonian can be constructed in the similar way as for the 2-body case. Eq.(\ref{3bodytransformation}) can be abbreviated as $\breve{R}_{i,i+1,i+2}(\eta,\beta,\varphi)|\psi(0,0,0)\rangle=|\psi(\eta,\beta,\varphi)\rangle$. On account of the Schr\"{o}dinger equation form $i\hbar\frac{\partial}{\partial t}|\psi\left(\eta,\beta,\varphi\right)\rangle=\hat{H}\left(\eta,\beta,\varphi\right)|\psi\left(\eta,\beta,\varphi\right)\rangle$ , the 3-body Hamiltonian takes the gauge potential form
$$\hat{H}_{i,i+1,i+2}=i\hbar\frac{\partial \breve{R}_{i,i+1,i+2}}{\partial t}\breve{R}_{i,i+1,i+2}^{-1}$$
that can be written explicitly as
\begin{eqnarray}\label{3bodyHamilton}
&&\hat{H}_{i,i+1,i+2}\nonumber\\
&=&\negthinspace\hbar\dot{\varphi}\{-\sin^2\eta\left[\tfrac{\sin^2\beta+1}{2}\left(S_{i}^3+S_{i+2}^3\right)+\cos^2\beta S_{i+1}^3\right]\nonumber\\
& &-\tfrac{1}{\sqrt{2}}\sin2\eta cos\beta\left(e^{i2\varphi}S_{i}^{+}S_{i+1}^{+}+e^{i2\varphi}S_{i+1}^{+}S_{i+2}^{+}+ h.c.\right)\nonumber\\
& &+\tfrac{1}{\sqrt{2}}\sin^2\eta \sin2\beta\left(S_{i}^{+}S_{i+1}^{-}+S_{i+1}^{+}S_{i+2}^{-}+ h.c.\right)\nonumber\\
& &-\sin2\eta \sin\beta\left(e^{i2\varphi}S_{i}^{+}S_{i+1}^{3}S_{i+2}^{+}+ h.c.\right)\nonumber\\
& &-\sin^2\eta \cos^2\beta\left(S_{i}^{+}S_{i+1}^{3}S_{i+2}^{-}+S_{i}^{-}S_{i+1}^{3}S_{i+2}^{+}\right)\}
\end{eqnarray}
where $S_{i}^{\pm}$ and $S_{i}^{3}$ represent the spin operators on i-th site and form $SU(2)$ algebra.

\medskip

The instantaneous eigenvalues and eigenstates of $\hat{H}_{i,i+1,i+2}$ are:

\medskip

 For $E_{0}=0$:
\begin{subequations}
\begin{align*}
 &|a_{1}\rangle=-\tfrac{1}{\sqrt{2}}\left(|011\rangle-|110\rangle\right)\nonumber\\
          &|a_{2}\rangle=-\tfrac{1}{\sqrt{2}}\left(|001\rangle-|100\rangle\right)\nonumber\\
          &|a_{3}\rangle=\tfrac{1}{\sqrt{1+\sin^2\beta}}\left(-\sqrt{2}\sin\beta|011\rangle+\cos\beta|101\rangle\right)\nonumber\\
          &|a_{4}\rangle=\tfrac{1}{\sqrt{1+\sin^2\beta}}\left(\sqrt{2}\sin\beta|001\rangle+\cos\beta|010\rangle\right)\nonumber
\end{align*}
\end{subequations}

For $E_{+}=2\hbar\dot{\varphi}\sin\eta$:
\begin{subequations}\label{Eplus}
\begin{align}
          |a_{5}\rangle=&-\tfrac{\cos\beta\sqrt{1-\sin\eta}}{2e^{-i2\varphi}}\left(|001\rangle-\sqrt{2}\tan\beta|010\rangle+|100\rangle\right)\nonumber\\
          &+\tfrac{\cos\eta}{\sqrt{2-2\sin\eta}}|111\rangle\\
          |a_{6}\rangle=&\tfrac{\cos\beta \cos\eta}{2\sqrt{1-\sin\eta}}(\tfrac{\sqrt{2}(\sin\eta-1)e^{i2\varphi}}{\cos\eta \cos\beta}|000\rangle\nonumber\\
          &+|011\rangle+\sqrt{2}\tan\beta|101\rangle+|110\rangle)
 \end{align}
\end{subequations}

 For $E_{-}=-2\hbar\dot{\varphi}\sin\eta$:
\begin{subequations}\label{Eminus}
\begin{align}
          |a_{7}\rangle=&\tfrac{\cos\beta\sqrt{1+\sin\eta}}{2e^{-i2\varphi}}(|001\rangle-\sqrt{2}\tan\beta|010\rangle+|100\rangle)\nonumber\\
          &+\tfrac{\cos\eta}{\sqrt{2+2\sin\eta}}|111\rangle\\
          |a_{8}\rangle=&\tfrac{\cos\beta \cos\eta}{2\sqrt{1+\sin\eta}}(\tfrac{\sqrt{2}(\sin\eta+1)e^{i2\varphi}}{\cos\eta  \cos\beta}|000\rangle+|011\rangle\nonumber\\
          &+\sqrt{2}\tan\beta|101\rangle+|110\rangle)\label{a8}
\end{align}
\end{subequations}
The states for $E_0=0$ are separable, whereas the other four states are entangled.

\medskip

With the eigenstates for the 3-body Hamiltonian, we can calculate the Berry phase in entanglement space. Referring  to Wilczek-Zee theory\cite{wilczek1984appearance,chrusscinnski2004geometric}, which is a natural generalization of Berry's theory, for degenerate spectra $H\psi_{na}=E_n\psi_{na}(a=1,2\ldots N)$, the corresponding adiabatic factor is $V=P \exp\left(i\oint_CA^{(n)}\right)$ with $A^{(n)}_{ab}:=i\langle\psi_{nb}|d|\psi_{na}\rangle$ and $C$ is a closed path in parameter space. But the eigenstates $|a_5\rangle\longrightarrow |a_8\rangle$ give that $A^{(n)}_{ab}=0$ for $a\neq b$, i.e. $A^{(n)}$ is diagonal. So it is much more convenient to express the geometric factor matrix as Berry phase\cite{berry1984quantal}.

When $\varphi\left(t\right)$ evolves a period adiabatically from $0$ to $\pi$ ($|a_1\rangle \longrightarrow |a_4\rangle$ are $\varphi$-independent, $|a_5\rangle \longrightarrow |a_8\rangle$ are $\varphi$-dependent), the Berry phases of the entangled states are:$$\gamma_{n}=i\int_{0}^{\pi}d\varphi\left\langle a_{n}\left|\tfrac{\partial}{\partial \varphi}\right|a_{n}\right\rangle \quad(n=5,6,7,8)$$
For $E_{+}=2\hbar\dot{\varphi}\sin\eta$:
\begin{subequations}
\begin{align}
\gamma_{5}=i\int_{0}^{\pi}d\varphi\left\langle a_{5}\left|\tfrac{\partial}{\partial \varphi}\right|a_{5}\right\rangle=-\pi(1-\sin\eta)\\
\gamma_{6}=i\int_{0}^{\pi}d\varphi\left\langle a_{6}\left|\tfrac{\partial}{\partial \varphi}\right|a_{6}\right\rangle=-\pi(1-\sin\eta)
\end{align}
\end{subequations}
For $E_{-}=-2\hbar\dot{\varphi}\sin\eta$:
\begin{subequations}
\begin{align}
\gamma_{7}=i\int_{0}^{\pi}d\varphi\left\langle a_{7}\left|\tfrac{\partial}{\partial \varphi}\right|a_{7}\right\rangle=-\pi(1+\sin\eta)\\
\gamma_{8}=i\int_{0}^{\pi}d\varphi\left\langle a_{8}\left|\tfrac{\partial}{\partial \varphi}\right|a_{8}\right\rangle=-\pi(1+\sin\eta)
\end{align}
\end{subequations}
Thus the Berry phases related to $\hat{H}_{i,i+1,i+2}$ are similar to the solid angle enclosed by the loop on the Bloch sphere depending on $\eta$ only. In Eq.(\ref{3Smatrix}), $\eta$ represents the rotation angle along the axis ``$\vec{n}\cdot\vec{\Sigma}$'', and $\beta$ represents the orientation of the axis.

\medskip

Now we discuss the eigenstates of $E_{\pm}$ in a little detail on the basis of Eq.(\ref{Eplus}) and (\ref{Eminus}).

\smallskip

(a) For the maximal energy gap $\left|E_{+}-E_{-}\right|$ at $\eta=\frac{\pi}{2}$. and by choosing $\sin\beta=-\frac{\sqrt{3}}{3}$, $\cos\beta=\frac{\sqrt{6}}{3}$ and $\varphi=0$, we have:
\begin{eqnarray*}
|a_{5}\rangle &=&|111\rangle \\
|a_{6}\rangle&=&\tfrac{1}{\sqrt{3}}\left(|011\rangle -|101\rangle+|110\rangle\right)\\[8pt]
|a_{7}\rangle &=&\tfrac{1}{\sqrt{3}}\left(|001\rangle +|010\rangle+|100\rangle\right) \\
|a_{8}\rangle&=&|000\rangle
\end{eqnarray*}
that are either direct product state or a genuine entangled W state.

\smallskip

(b) As $\eta=\frac{\pi}{6}$, $\sin\beta=-\frac{\sqrt{3}}{3}$, $\cos\beta=\frac{\sqrt{6}}{3}$ and $\varphi=0$, we have:
\begin{eqnarray*}
|a_{5}\rangle &=&-\tfrac{\sqrt{3}}{6}\left(|001\rangle+|010\rangle+|100\rangle-3|111\rangle\right) \\
|a_{6}\rangle &=&-\tfrac{1}{2}\left(|000\rangle-|011\rangle+|101\rangle-|110\rangle\right) \\[8pt]
|a_{7}\rangle &=&\tfrac{1}{2}\left(|001\rangle+|010\rangle+|100\rangle+|111\rangle\right)  \\
|a_{8}\rangle &=&\tfrac{\sqrt{3}}{6}\left(3|000\rangle+|011\rangle-|101\rangle+|110\rangle\right)
\end{eqnarray*}
here $|a_{6}\rangle$ and $|a_{7}\rangle$ are GHZ states under local unitary transformation. When $\eta=-\frac{\pi}{6}$, $|a_{5}\rangle$ and $|a_{8}\rangle$ are GHZ states under local unitary transformation.

\bigskip

\section{Next neighbour spin-1/2 chain model for 3-body S-matrix and 1D Kitaev toy model}
\label{sec:4}

Majorana fermions(MFs) are particles that are their own anti-particles and obey non-Abelian statistics\cite{ivanov2001non,alicea2011non,stern2010non}, which attract much attention due to their potential application in topological quantum computation\cite{nayak2008non,leijnse2012introduction}. In Ref.\cite{kitaev2001unpaired}, Kitaev proposed a spinless chain model: a ``quantum wire'' lies on the surface of three-dimensional superconductor. This model generates unpaired Majorana fermions(topological ground state degeneracy). The chain consists of $L\gg1$ sites, with each site being either empty or occupied by an electron(with a fixed spin direction). The Hamiltonian of the toy model reads\cite{kitaev2001unpaired}:
\begin{eqnarray}\label{ToyModel}
\hat{H}_k = \sum_j^L &&[-\mu\left(a_j^{\dag}a_{j}-\tfrac{1}{2}\right)-\omega\left(a_j^{\dag}a_{j+1}+a_{j+1}^{\dag}a_{j}\right)\nonumber\\
&&+\Delta a_ja_{j+1}+\Delta^{\ast}a_{j+1}^{\dag}a_{j}^{\dag}]
\end{eqnarray}
here $\omega$ is hopping amplitude, $\mu$ is chemical potential, $\Delta=|\Delta|e^{-i2\varphi}$ is induced superconducting gap. By defining Majorana fermion operators:
\begin{subequations}
\begin{align}\label{MFs}
 c_{2j-1}&=e^{i\varphi}a_j^{\dag}+e^{-i\varphi}a_j\\
 c_{2j}&=i e^{i\varphi}a_j^{\dag}-i e^{-i\varphi}a_j
 \end{align}
\end{subequations}
They satisfy the relations:
\begin{equation}
c_{m}^{\dag}=c_{m}, \quad c_{l}c_{m}+c_{m}c_{l}=2\delta_{lm}, \quad l,m=1,\ldots 2N
\end{equation}
Then the Hamiltonian is transformed to be:
\begin{eqnarray}
\hat{H}_k=\tfrac{i}{2}\sum_j&& [-\mu c_{2j-1}c_{2j}+\left(\omega+|\Delta|\right)c_{2j}c_{2j+1}\nonumber\\
&&+\left(-\omega+|\Delta|\right)c_{2j-1}c_{2j+2}]
\end{eqnarray}

Now let us compare our chain models(including 2-body S-matrix and 3-body S-matrix) with 1D Kitaev toy model.
We first consider the chain for 2-body S-matrix $\breve{R}_{i,i+1}\left(\theta,\varphi\right)$. Recalling the Hamiltonian $\hat{H}_{i,i+1}=i\hbar \frac{\partial \breve{R}}{\partial t}\breve{R}^{\dag}$, then the form:
\begin{eqnarray}\label{twobodyHamilton}
 \hat{H}_{i,i+1}&=&-\hbar \dot{\varphi}\sin\theta [\tfrac{\sin\theta}{2}\left(S_i^3+S_{i+1}^3\right)\nonumber\\
 &&+\cos\theta\left(e^{i\varphi}S_i^+S_{i+1}^++e^{-i\varphi}S_i^-S_{i+1}^-\right)]
\end{eqnarray}
where $S_i^+=\left(\begin{array}{cc}
  0 & 1 \\
  0 & 0
  \end{array}\right)
$, $S_i^-=\left(\begin{array}{cc}
  0 & 0 \\
  1 & 0
  \end{array}\right)$, $S_i^3=\left(\begin{array}{cc}
  1 & 0 \\
  0 & -1
  \end{array}\right)$ are spin operators at i-th site.

\medskip
Making the Jordan-Wigner transformation to represent spin operators at sites with spinless fermion operators, we have:
\begin{eqnarray}\label{JWtransformation}
 S_j^+&=&e^{-i\pi\Sigma_{k=1}^{j-1}a_k^{\dag}a_k}a_j^{\dag}\nonumber\\
 S_j^-&=&e^{i\pi\Sigma_{k=1}^{j-1}a_k^{\dag}a_k}a_j\\
 S_j^3&=&2a_j^{\dag}a_j-1\nonumber
\end{eqnarray}
The chain model Hamiltonian reads:
\begin{eqnarray}
\hat{H}_2&=&\sum_{i=1}^{N}\hat{H}_{i,i+1}\label{2bodyChain}\\
\hat{H}_{i,i+1}&=&-\hbar\dot{\varphi}\sin\theta[\sin\theta(a_i^{\dag}a_i+a_{i+1}^{\dag}a_{i+1}-1)\nonumber\\
& &+\cos\theta(e^{i\varphi}a_{i+1}^{\dag}a_i^{\dag}+e^{-i\varphi}a_{i}a_{i+1})]\label{2bodyspinless}
\end{eqnarray}
In comparison with the 1D Kitaev toy model, the chain model for 2-body S-matrix $\hat{H}_2$ is a special case of 1D Kitaev toy model without hopping term($\omega=0$). But in chain model for 3-body S-matrix Hamiltonian, there appears hopping term that we shall show below.

\medskip

Denoting $\vec{S}_i^3=(S_i^3+S_{i+1}^3)/2$, $\vec{S}_i^+=S_i^+S_{i+1}^+$ and $\vec{S}_i^-=S_i^-S_{i+1}^-$, they still form $SU(2)$ algebra with spin $\frac{1}{2}$. And then Eq.(\ref{twobodyHamilton}) can be written in the following form:
\begin{eqnarray}
 \hat{H}_{i,i+1} &=& -\hbar \dot{\varphi}\sin\theta \left[\sin\theta S^z+\cos\theta \cos\varphi S^x+\cos\theta \sin\varphi S^y\right]\nonumber\\
   &=& (-\hbar\omega)\cos\alpha \overrightarrow{n}\cdot\overrightarrow{S}
\end{eqnarray}
where
\begin{eqnarray*}
  \overrightarrow{S} &=& (S^{x},S^{y},S^{z})\\
  \overrightarrow{n} &=& (\cos\theta \cos\varphi,\cos\theta \sin\varphi,\sin\theta)
\end{eqnarray*}
It is easy to be diagonalized by unitary rotation. Therefore, Kitaev's model without hopping term turns into a Nuclear-Magnetic-Resonance(NMR) problem of a ``bispin'' system that comes from Type-II solution of YBE.

\medskip

In Sec.\ref{Sec:3}, the Hamiltonian for 3-body  $\hat{H}_{i,i+1,i+2}$ (see Eq.\ref{3bodyHamilton}) has been obtained. Now we recast Eq.(\ref{3bodyHamilton}) to homogeneous chain model, for $\eta$, $\beta$ and $\varphi$ are same in different sites. Then the total spin-$\frac{1}{2}$ chain model is:
\begin{equation}\label{3bodyChain}
\hat{H}_{3}=\sum_{n=1}^{N} \hat{H}_{n,n+1,n+2}
\end{equation}
In comparison with Eq.(\ref{2bodyChain}), the Eq.(\ref{3bodyChain}) contains not only the nearest neighbour interaction for $(n, n+1)$-th sites and $(n+1, n+2)$-th sites, but also the next nearest neighbouring interaction for $(n,n+2)$-th sites.

\medskip

Now we compare the chain model for 3-body S-matrix with 1D Kitaev's toy model\cite{kitaev2001unpaired}.

\medskip

After Jordan-Wigner transformation(Eq.\ref{JWtransformation}), the Hamiltonian(Eq.\ref{3bodyChain}) turns out to be($\hbar\dot{\varphi}=1$ for simplicity):
\begin{subequations}\label{3bodyToymodel}
\begin{align}
\hat{H}_3
=&\sum_{n=1}^{N-2}[-\sin^2\eta(1+\sin^2\beta)(a_{n}^{\dag}a_{n}-\tfrac{1}{2})\nonumber\\
& -\tfrac{\sqrt{2}}{2}\sin2\eta \cos\beta(e^{i2\varphi} a_{n}^{\dag}a_{n+1}^{\dag}+h.c.)\nonumber\\
& +\tfrac{\sqrt{2}}{2}\sin^2\eta \sin2\beta(a_{n}^{\dag}a_{n+1}-a_{n}a_{n+1}^{\dag})]\label{first}\\
& +\sum_{n=1}^{N-2}[-2\sin^2\eta \cos^2\beta(a_{n+1}^{\dag}a_{n+1}-\tfrac{1}{2})\nonumber\\
& -\tfrac{\sqrt{2}}{2}\sin2\eta \cos\beta(e^{i2\varphi} a_{n+1}^{\dag}a_{n+2}^{\dag}+h.c.)\nonumber\\
& +\tfrac{\sqrt{2}}{2}\sin^2\eta \sin2\beta(a_{n+1}^{\dag}a_{n+2}-a_{n+1}a_{n+2}^{\dag})]\label{second}\\
& +\sum_{n=1}^{N-2}[-\sin^2\eta(1+\sin^2\beta)(a_{n+2}^{\dag}a_{n+2}\nonumber\\
& -\tfrac{1}{2})+\sin2\eta \sin\beta(e^{i2\varphi} a_{n}^{\dag}a_{n+2}^{\dag}+ h.c.)\nonumber\\
& +\sin^2\eta \cos^{2}\beta(a_{n}^{\dag}a_{n+2}-a_{n}a_{n+2}^{\dag})]\label{third}
\end{align}
\end{subequations}
We can see that $\hat{H}_3$ is obviously the extension of 1D Kitaev toy model(Eq.\ref{ToyModel}). There are three parts in the chain $\hat{H}_3$. The first part(Eq.\ref{first}) includes fermions from 1 to N-1 sites, the second part(Eq.\ref{second}) includes fermions from 2 to N sites. Both of them form the Kitaev model with specified parameters. The third part(Eq.\ref{third}) includes fermions from 3 to N sites and gives rise to the interaction terms between $n$ and $n+2$ sites.
Suppose $N\gg 1$, after Fourier transformation:
\begin{eqnarray}
a_{n}=&\sqrt{\frac{1}{N}}\sum_{k=-\pi}^{\pi}e^{-ink}f_{k}\nonumber\\
a_{n}^{\dag}=&\sqrt{\frac{1}{N}}\sum_{k=-\pi}^{\pi}e^{ink}f_{k}^{\dag}
\end{eqnarray}
The Hamiltonian is transformed into:
\begin{eqnarray}
\hat{H} &=& \sum_{k=-\pi}^{\pi}\left(X_{k}S_{k}^{z}+Y_{k}S_{k}^{+}+Y_{k}^{*}S_{k}^{-}\right)\\
X_{k} &=& -\sin^2\eta\left[6-(\sqrt{2}\sin\beta+2\cos\beta \cos k)^2\right]\nonumber\\
Y_{k} &=& i\sin2\eta(\sqrt{2}\cos\beta \sin k-\sin\beta \sin2k)e^{i2\varphi}\nonumber\\
Y_{k}^{*} &=& -i\sin2\eta(\sqrt{2}\cos\beta \sin k-\sin\beta \sin2k)e^{-i2\varphi}\nonumber
\end{eqnarray}
where $S_{k}^{z}=\frac{1}{2}(f_{k}^{\dag}f_{k}+f_{-k}^{\dag}f_{-k}-1)$, $S_{k}^{+}=f_{k}^{\dag}f_{-k}^{\dag}$ and $S_{k}^{-}=f_{-k}f_{k}$  form $SU(2)$ algebra. The eigenspectra (energy vs. momentum) contains two bands of quasiparticle excitations:
\begin{equation}\label{energygap}
\epsilon \left(k\right)=\pm\sqrt{\tfrac{|X_{k}^2|}{4}+|Y_{k}|^2}
\end{equation}

\medskip

1D Kitaev toy model is a model with $\mathbb{Z}_2$ symmetry and topological degeneracy. The appearance of ground state degeneracy is dependent on the boundary condition of the chain. The ground state is 2-fold degeneracy for an open chain and unique on a closed loop\cite{kitaev2001unpaired,kitaev2009topological}.
The basic idea of Kitaev's model is the existence of ``Zero Mode'', which makes the appearance of unpaired Majorana fermions at the end of the chain possible. The generalized condition for generating unpaired Majorana fermions is given by\cite{kitaev2001unpaired}: $2|\omega|>|\mu|$(see Eq.\ref{ToyModel}), and $2|\omega|=|\mu|$ is a phase boundary (in this boundary the bulk energy gap vanishes). We now investigate whether the existence of the so called ``Zero Mode'' in our chain model for 3-body S-matrix. Our chain model can be written in terms of Majorana operators $c_{j}$(see Eq.\ref{MFs}):
\begin{eqnarray}\label{3BChainMF}
&&-2i\hat{H}_{3}\nonumber\\
&=&-\mu_{1}(c_{1}c_{2}+c_{2N-1}c_{2N})-(\mu_{1}+\mu_{2})(c_{3}c_{4}+c_{2N-3}c_{2N-2})\nonumber\\
&&+(\Delta_{1}+\omega_{1})(c_{2}c_{3}+2c_{4}c_{5}+c_{2N-2}c_{2N-1}+2c_{2N-4}c_{2N-3})\nonumber\\
&&+(\Delta_{1}-\omega_{1})(c_{1}c_{4}+2c_{3}c_{6}+c_{2N-3}c_{2N}+2c_{2N-5}c_{2N-2})\nonumber\\
&&+(\Delta_{2}+\omega_{2})(c_{2}c_{5}+c_{4}c_{7}+c_{2N-4}c_{2N-1}+c_{2N-6}c_{2N-3})\nonumber\\
&&+(\Delta_{2}-\omega_{2})(c_{1}c_{6}+c_{3}c_{8}+c_{2N-5}c_{2N}+c_{2N-7}c_{2N-2})\nonumber\\
&&+\sum_{j=3}^{N-2}(-2\mu_{1}-\mu_{2})c_{2j-1}c_{2j}\nonumber\\
&&+\sum_{j=3}^{N-3}[2(\omega_{1}+\Delta_{1})c_{2j}c_{2j+1}+2(-\omega_{1}+\Delta_{1})c_{2j-1}c_{2j+2}\nonumber\\
&&+(\omega_{2}+\Delta_{2})c_{2j}c_{2j+3}+(-\omega_{2}+\Delta_{2})c_{2j-1}c_{2j+4}]
\end{eqnarray}
with the parameter:
\begin{subequations}\label{3Bparameter}
\begin{align}
&\mu_{1}=\sin^2\eta(1+\sin^2\beta) \\ &\mu_{2}=2\sin^2\eta \cos^2\beta\\
&\omega_{1}=-\sqrt{2}\sin^2\eta \cos\beta \sin\beta \\ &\omega_{2}=-\sin^2\eta \cos^2\beta\\
&\Delta_{1}=-\sqrt{2}\sin\eta \cos\eta \cos\beta \\ &\Delta_2=2\sin\eta \cos\eta \sin\beta
\end{align}
\end{subequations}
Then the Hamiltonian(Eq.\ref{3BChainMF}) takes the form:
\begin{equation}
\hat{H}_{3}=\frac{i}{4}\sum_{l,m}A_{lm}c_{l}c_{m}, \quad A_{lm}^{\ast}=A_{lm}=-A_{ml}
\end{equation}
which can reduce to
\begin{equation}
\hat{H}_{3}=\frac{i}{2}\sum_{m=1}^{N}\epsilon_{m}b_{m}'b_{m}''
\end{equation}
where $b_{m}'$ and $b_{m}''$ are real linear combinations of $c_{2j-1}$, $c_{2j}$ with the same commutation relations. ``Zero Mode'' means $\epsilon_{m}=0$ for some $m$. For the simplicity we suppose $\omega_{2}=\Delta_{2}$\footnote{Ref.\cite{kitaev2001unpaired} has shown that the existence condition of unpaired MFs(phase transition) is the hopping and chemical potential parameter $2|\omega|>|\mu|$, $|\Delta|$ is related to the bulk gap. In Eq.(\ref{3bodyToymodel}), the corresponding hopping and chemical potential parameter have the same factor $\sin^2\eta$, so $\eta$ does not influence the existence of unpaired MFs($\eta\neq0$).}, the mode becomes the form(see Appendix B):
\begin{subequations}
\begin{align}
&b'=\sum_{j=1}^{N}\left(\alpha_{1}'x_{1}^{j}+\alpha_{2}'x_{2}^{j}+\alpha_{3}'x_{3}^{j}\right)c_{2j}\\
&b''=\sum_{j=1}^{N}\left(\alpha_{1}''x_{1}^{-j}+\alpha_{2}''x_{2}^{-j}+\alpha_{3}''x_{3}^{-j}\right)c_{2j-1}
\end{align}
\end{subequations}
where $x_{1}$, $x_{2}$ and $x_{3}$ are roots of equation $(2\tan^2\beta+1)x^3-2\sqrt{2}(\tan^3\beta+\tan\beta)x^2+(2\tan^2\beta-1)x+\sqrt{2}\tan\beta=0$ obtained from the Hamiltonian(Eq.\ref{3BChainMF}). $b'$ and $b''$ represent two unpaired Majorana fermions at the end of the chain.
\begin{figure}[!ht]
\centering
\includegraphics[scale=0.4]{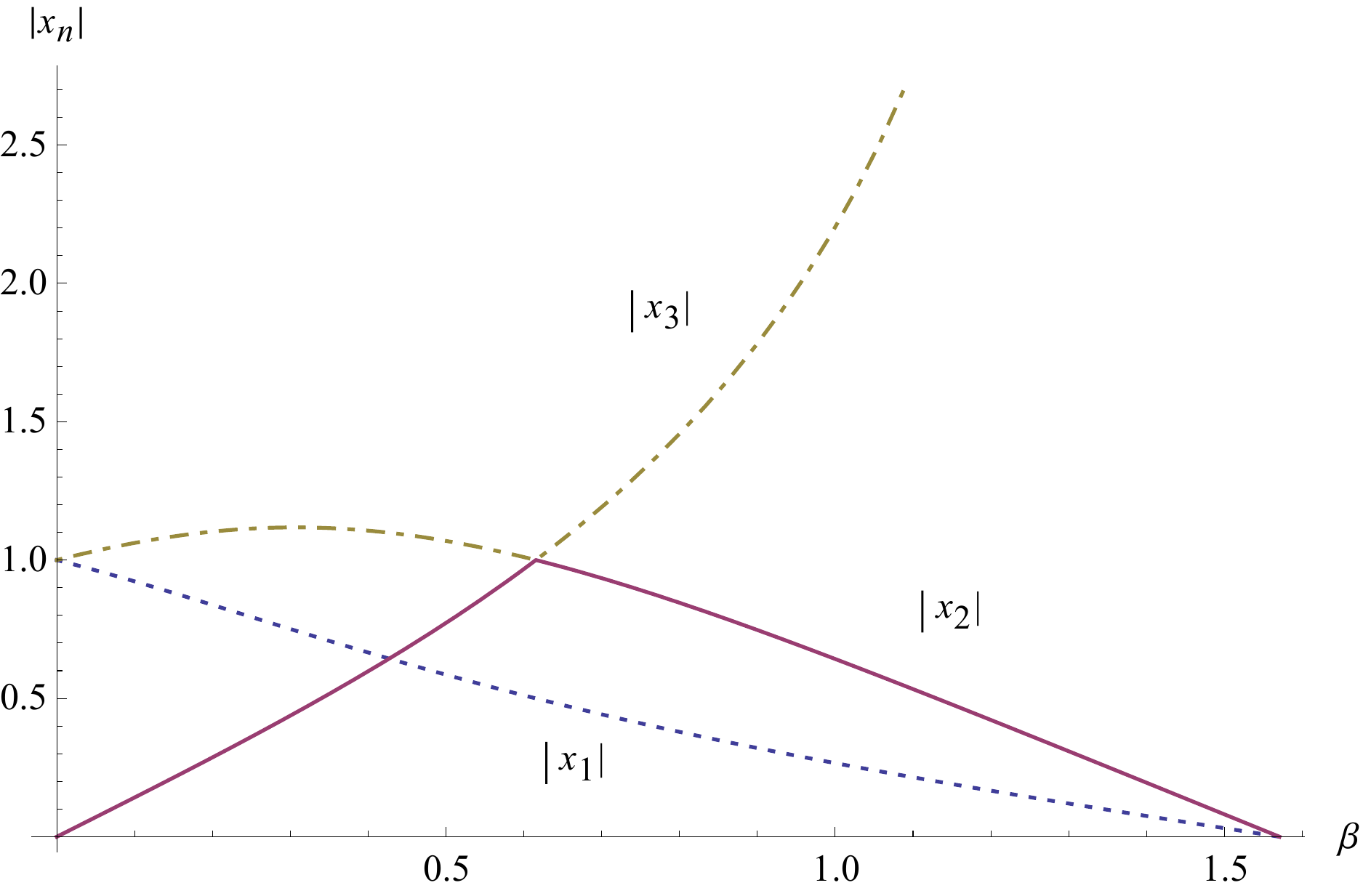}
\caption{ Modulus of three roots $x_1$, $x_2$ and $x_3$.  The dashed line represents the modulus of $x_1$, the solid line represents the modulus of $x_2$, the dotdashed line represents the modulus of $x_3$. we find that when $\beta=\arccos(\frac{\sqrt{6}}{3})$, $x_1=0.5$, $|x_2|=|x_3|=1$, and the energy gap disappears, which is similar to the ``phase boundary'' $2|\omega|=|\mu|$ proposed in 1D Kitaev toy model.}\label{Figure1}
\end{figure}

\medskip

For $0\leq\beta<\frac{\pi}{2}$, by numerical calculation we can find two roots $|x_{1}|\leq 1$ , $|x_{2}|\leq 1$ and the third one $|x_{3}|\geq1$(See Fig.\ref{Figure1}), therefore, to generate unpaired Majorana fermions at the end of the chain under the boundary condition, $\alpha_{3}'$ and $\alpha_{3}''$ should be zero. However, we shall show that $\alpha_{1}'$ and $\alpha_{2}'$($\alpha_{1}''$ and $\alpha_{2}''$) cannot be zero under the boundary condition. The boundary condition reads(see the matrix in Appendix B):
\begin{subequations}
\begin{align}
&\sum_{i=1}^{2}\alpha_{i}'[-\mu_{1}x_i+(-\omega_{1}+\Delta_{1})x_{i}^{2}]=0\label{boundCon1}\\
&\sum_{i=1}^{2}\alpha_{i}'[-\omega_{2}-\Delta_{2}+(-\omega_{1}-\Delta_{1})x_{i}-\mu_{1}x_i^2]=0\label{boundCon2}\\
&\sum_{i=1}^{2}\alpha_{i}''x_{i}^{-N-1}[\mu_{1}x_i+(\omega_{1}-\Delta_{1})x_{i}^2]=0\label{boundCon3}\\
&\sum_{i=1}^{2}\alpha_{i}''x_{i}^{-N-1}[-\omega_{2}-\Delta_{2}+(-\omega_{1}-\Delta_{1})x_{i}-\mu_{1}x_i^{2}]=0\label{boundCon4}
\end{align}
\end{subequations}
 The existence of topological ground state degeneracy depends only on the parameter $\beta$. Now we consider two cases: 1) $\beta\neq \arccos\frac{\sqrt{6}}{3}$, $\alpha_{1}'=\alpha_{2}'=\alpha_{1}''=\alpha_{2}''=0$, which means that there are not unpaired Majorana fermions at the end of the chain; 2) $\beta=\arccos\frac{\sqrt{6}}{3}$,  the bulk energy gap between ground state and excited state disappears for the root $|x_1|=0.5$, $|x_2|=1$, $|x_3|=1$, and it is impossible to generate unpaired Majorana fermions near the end of the chain. Thus we conclude that the unpaired Majorana fermions are killed by YBE. Similar to the ``phase boundary'' mentioned in the Ref.\cite{kitaev2001unpaired}, the bulk gap in our chain disappears when $\eta=-\frac{\pi}{3}$ and $\beta=\arccos\frac{\sqrt{6}}{3}$, although there is not phase transition as Kitaev's model in our chain. In the meanwhile, $\beta=\arccos\frac{\sqrt{6}}{3}$ and $\eta=-\frac{\pi}{3}$ are a sufficient condition for generating 3-qubit GHZ state in Sec.\ref{Sec:2}.

  Let us give a brief discussion why YBE does not allow unpaired MFs in Eq.(\ref{ToyModel}). According to the Ref.\cite{kitaev2001unpaired}, the existence condition of unpaired Majorana fermions is $2|\omega|>|\mu|$. Eq.(\ref{3bodyToymodel}) shows that each part (\ref{first}, \ref{second}, \ref{third}) can be viewed as a Kitaev model. If we suppose each part has unpaired MFs, the condition should be
 \begin{subequations}
 \begin{align}
 &2|\tfrac{\sqrt{2}}{2}\sin^2\eta\sin2\beta|>|\sin^2\eta(1+\sin^2\beta)|\label{Confirst}\\
 &2|\tfrac{\sqrt{2}}{2}\sin^2\eta\sin2\beta|>|2\sin^2\eta\cos^2\beta|\label{Consecond}\\
 &2|\sin^2\eta\cos^2\beta|>|\sin^2\eta(1+\sin^2\beta)|\label{Conthird}
 \end{align}
 \end{subequations}
 If one of the three conditions(\ref{Confirst},\ref{Consecond},\ref{Conthird}) is satisfied, another two conditions are violated, so it is impossible to generate unpaired MFs. The reason is that YBE has set the parameters in Eq.(\ref{ToyModel}) to be dependent. The allowed region is out of the condition for the existence of unpaired MFs.

\section{Entangling dependence on parameter  in special chain model }
\label{Sec:5}

In Sec.{\ref{Sec:2}}, we show that the condition satisfied by parameter $\eta$ for generating GHZ states and W states are $\eta=\frac{\pi}{3}$ and $\frac{\pi}{2}$, respectively. Now we discuss a special case of the chain model for 3-body S-matrix . Two issues will be taken into account below. Firstly, we discuss the entanglement transfer in our chain model by adding Aharonov-Casher effect on spin sites; Secondly, we discuss the example for the entanglement transfer dependence on entanglement space parameter $\beta$ for N=4.
\medskip

 The condition $\eta=\frac{\pi}{2}$ corresponds to both W states and the maximal energy gap between $E_{\pm}$ in the Hamiltonian for 3-body S-matrix (Sec.\ref{Sec:3}). The chain model(Eq.\ref{3bodyChain}) then becomes($\hbar=\dot{\varphi}=1$):
\begin{eqnarray}\label{Spe3BChain}
 \hat{H}&=&-\sum_{n=1}^{N}\left[\tfrac{1+\sin^2\beta}{2}\left(S_{n}^{3}+S_{n+2}^{3}\right)+\cos^2\beta S_{n+1}^{3}\right]\nonumber\\
 &&+\tfrac{\sqrt{2}}{2}\sin2\beta\sum_{n=1}^{N}\left(S_{n}^{+}S_{n+1}^{-}+S_{n+1}^{+}S_{n+2}^{-}+ h.c.\right)\nonumber\\
&&-\cos^2\beta\sum_{n=1}^{N}\left(S_{n}^{+}S_{n+1}^{3}S_{n+2}^{-}+S_{n}^{-}S_{n+1}^{3}S_{n+2}^{+}\right)
\end{eqnarray}
By choosing periodic boundary condition(N+1=1,N+2=2), the above equation turns out to be
\begin{eqnarray}
\hat{H}&=&-\sum_{n=1}^{N}2S_{n}^{3}+\sqrt{2}\sin2\beta \sum_{n=1}^{N}\left(S_{n}^{+}S_{n+1}^{-}+S_{n}^{-}S_{n+1}^{+}\right)\nonumber\\
 &&-\cos^2\beta\sum_{n=1}^{N}\left(S_{n}^{+}S_{n+1}^{3}S_{n+2}^{-}+S_{n}^{-}S_{n+1}^{3}S_{n+2}^{+}\right)
\end{eqnarray}
After Jordan-Wigner transformation(see Eq.\ref{JWtransformation}) the transformed chain model becomes:
\begin{eqnarray}\label{Specfermion}
\hat{H}&=&-2\sum_{n=1}^{N}\left(2a_{n}^{\dag}a_{n}-1\right)+\cos^2\beta\sum_{n=1}^{N-2}\left(a_{n}^{\dag}a_{n+2}-a_{n}a_{n+2}^{\dag}\right)\nonumber\\
 &&+\sqrt{2}\sin2\beta \sum_{n=1}^{N-1}\left(a_{n}^{\dag}a_{n+1}-a_{n}a_{n+1}^{\dag}\right)\nonumber\\
&&-\sqrt{2}\sin^2\beta \left[a_{N}^{\dag}a_{1}-a_{N}a_{1}^{\dag}\right]e^{i\pi\sum_{k=1}^{N}a_{k}^{\dag}a_{k}}\nonumber\\
 &&-\cos^2\beta\left(a_{N-1}^{\dag}a_{1}+a_{N}^{\dag}a_{2}\right)e^{-i\pi\sum_{k=1}^{N}a_{k}^{\dag}a_{k}}\nonumber\\
 &&+\cos^2\beta\left(a_{N-1}a_{1}^{\dag}+a_{N}a_{2}^{\dag}\right)e^{i\pi\sum_{k=1}^{N}a_{k}^{\dag}a_{k}}
\end{eqnarray}
To solve this model, we follow Ref.\cite{maruyama2007enhancement} by assuming that the system is in the ``one-magnon'' state, namely the total number of spin flip in the chain model is one. Under the condition, $e^{i\pi\sum_{k=1}^{N}a_{k}^{\dag}a_{k}}=e^{-i\pi\sum_{k=1}^{N}a_{k}^{\dag}a_{k}}=-1$. Through Fourier transformation:
\begin{eqnarray}
a_{k}^{\dag}&=&\tfrac{1}{\sqrt{N}}\sum_{j}e^{i2\pi kj/N}\eta_{j}^{\dag}\\
a_{k}&=&\tfrac{1}{\sqrt{N}}\sum_{j}e^{-i2\pi kj/N}\eta_{j}\nonumber
\end{eqnarray}
where  $1\leq j\leq N$. The Eq.(\ref{Specfermion}) is transformed into
\begin{equation}
\hat{H}=\sum_{j}\left(E_{j}\eta_{j}^{\dag}\eta_{j}+2\right)
\end{equation}
The energy spectrum(ignore the constant term 2N) is
\begin{equation}\label{eigenenergy}
E_{j}=2\sqrt{2}\sin2\beta \cos(\tfrac{2\pi j}{N})+2\cos^{2}\beta \cos(\tfrac{4\pi j}{N})-4
\end{equation}
and the corresponding eigenstate
\begin{equation}\label{FourierEigen}
|j\rangle:=\eta_{j}^{\dag}|\downarrow\rangle^{\otimes N}=\tfrac{1}{\sqrt{N}}\sum_{k=1}^{N}e^{i2\pi \tfrac{kj}{N}}S_{k}^{+}\left|\downarrow\right\rangle^{\otimes N}
\end{equation}

\medskip

In Ref.\cite{maruyama2007enhancement}, a description of the entanglement transfer between an arbitrary pair of spins for an N-spin chain under the one-magnon condition has been given:
\begin{equation}\label{OneMagnon}
\left|\Psi(t)\right\rangle=\sum_{k=1}^{N}\alpha_{k}(t)S_{k}^{+}|0\rangle^{\otimes N}
\end{equation}
where $S_{m}^{+}=S_{m}^{x}+iS_{m}^{y}$ is the raising operator defined with the spin-$\frac{1}{2}$ operators $S_{m}^{\alpha}(\alpha=x,y,z)$ for the m-th spin. Following Ref.\cite{maruyama2007enhancement} we are going to discuss how to measure the pairwise entanglement in $|\Psi(t)\rangle$ by introducing the concurrence. The concurrence $C$ in a bipartite reduced density matrix $\rho$ is defined as\cite{wootters1998entanglement}:
\begin{equation}
C:=\max\{0,\lambda_{1}-\lambda_{2}-\lambda_{3}-\lambda_{4}\}
\end{equation}
where $\lambda_{i}$ are the square roots of eigenvalues of the matrix $R$ in descending order. The matrix $R$ is given as a product of $\rho$ and its time-reversed state, namely
\begin{equation}
R=\rho\left(\sigma^{y}\otimes\sigma^{y}\right)\rho^{*}\left(\sigma^{y}\otimes\sigma^{y}\right)
\end{equation}
the concurrence takes its maximum value 1 for the maximal entangled state, and 0 for all separable states. For the concurrence between $l_1$th and $l_2$th spins at time $t$, the density matrix $\rho(l_{1},l_{2})$ can be evaluted by tracing out all spins except these two. The concurrence $C_{l_{1},l_{2}}$ is expressed as
\begin{equation}
C_{l_{1},l_{2}}(t)=2|\alpha_{l_{1}}(t)||\alpha_{l_{2}}(t)|
\end{equation}

\medskip

Now turn back to our chain model. Suppose at time $t=0$ the $m_{1}$th spin and the $m_{2}$th spin are maximally entangled, then the state can be expressed as
\begin{equation}
\left|\Psi(0)\right\rangle=\tfrac{1}{\sqrt{2N}}\sum_{j}\left[e^{-i2\pi \tfrac{jm_{1}}{N}}+e^{-i2\pi \tfrac{jm_{2}}{N}}\right]|j\rangle
\end{equation}
If the state evolves adiabatically, there should be an extra dynamical phase for each state $|j\rangle$($\hbar=1$) at time $t$:
\begin{equation}\label{initialstate}
\left|\Psi(t)\right\rangle=\tfrac{1}{\sqrt{2N}}\sum_{j}\left(e^{-i2\pi
\tfrac{jm_{1}}{N}}+e^{-i2\pi \tfrac{jm_{2}}{N}}\right)e^{-iE_{j}t}|j\rangle
\end{equation}
Substitute Eq.(\ref{FourierEigen}) into Eq.(\ref{initialstate}), it leads to

\begin{eqnarray}
|\Psi(t)\rangle\nonumber&=&\tfrac{1}{\sqrt{2}N}\sum_{k,j}(e^{i2\pi \tfrac{j(k-m_{1})}{N}-iE_{j}t}\nonumber\\
&&+e^{i2\pi \tfrac{j(k-m_{2})}{N}-iE_{j}t})S_{k}^{+}| \downarrow \rangle^{ \otimes N}
\end{eqnarray}
In comparison with Eq.(\ref{OneMagnon}), we get the formula of $\alpha_{j}(t)$:$(1 \leq j \leq N)$
\begin{eqnarray}
&&\alpha_{k}^{m_{1},m_{2}}(t)\nonumber\\
&=&\tfrac{1}{\sqrt{2}N}\sum_{j}\left(e^{i2\pi \tfrac{j(k-m_{1})}{N}}+e^{i2\pi \tfrac{j(k-m_{2})}{N}}\right)e^{-iE_{j}t}
\end{eqnarray}

With the obtained $\alpha_{k}(t)$, we evaluate the concurrence between $l_{1}$th and $l_{2}$th spin at time $t$:
\begin{equation}\label{TransCon}
C_{l_{1},l_{2}}^{m_{1},m_{2}}=2|\alpha_{l_{1}}^{m_{1},m_{2}}(t)||\alpha_{l_{2}}^{m_{1},m_{2}}(t)|
\end{equation}

\medskip

We first detect the influence of adding Aharonov-Casher phase in our chain model. In Ref.\cite{maruyama2007enhancement}, the effect of entanglement transfer by adding AC phase on standard Heisenberg model that has only nearest neighbouring interaction has been discussed, but in our chain model both nearest neighbouring and next neighbouring interactions are included.

\medskip

As is known well, the Aharonov-Casher effect is proposed in Ref.\cite{aharonov1984topological}, and can be taken as a physical mechanism to cause a phase shift\cite{maruyama2007enhancement,cao1997quantum,meier2003magnetization}. For a neutral particle with magnetic moment $\vec{\mu}$ travels from $\vec{r}$ to $(\vec{r}+\Delta\vec{r})$ in the electric field $\vec{E}$, the wave function of particle acquires  an extra phase, A-C phase:
\begin{equation}\label{ACphase}
\Delta\theta=\frac{1}{\hbar c^2}\int_{\vec{r}}^{\vec{r}+\Delta\vec{r}}\vec{\mu}\times\vec{E}(x)\cdot d\vec{x}
\end{equation}

Consider our chain model(Eq.\ref{Spe3BChain}) as a ring-shaped chain, i.e. with the periodic boundary condition $N+1=1, N+2=2$. The phase change $\theta$ between $i$ and $i+1$-th spin site is given by Eq.(\ref{ACphase}) with $\vec{r}=\vec{r}_i$ and $\Delta\vec{r}=\vec{r}_{i+1}-\vec{r}_{i}$, where $2\theta$ is the phase change between $i$ and $i+2$-th spin site. Now under the applied field Eq.(\ref{Spe3BChain}) is transformed into:
\begin{eqnarray}
\hat{H}&=&-\sum_{n=1}^{N}2S_{n}^{3}+\sqrt{2}\sin2\beta \sum_{n=1}^{N}\left(e^{i\theta}S_{n}^{+}S_{n+1}^{-}+h.c.\right)\nonumber\\
 &&-\cos^2\beta\sum_{n=1}^{N}\left(e^{i2\theta}S_{n}^{+}S_{n+1}^{3}S_{n+2}^{-}+h.c.\right)
\end{eqnarray}

The  eigenenergy after adding AC phase is
\begin{eqnarray}
E_{j}&=&-4+2\sqrt{2}\sin 2\beta \cos(\theta-2\pi j/N)\nonumber\\
&&+2\cos^{2}\beta \cos(2\theta-4\pi j/N)
\end{eqnarray}

\begin{figure}[!ht]
\centering
\includegraphics[scale=0.45]{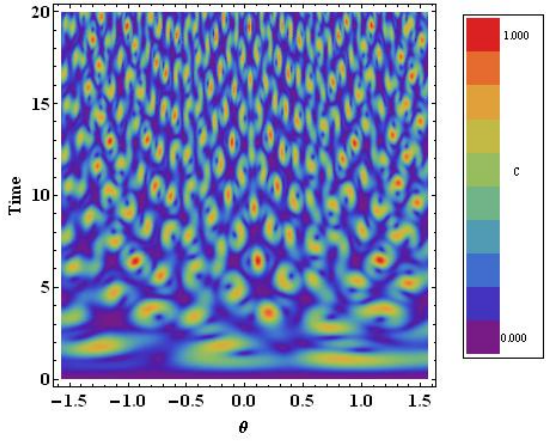}
\caption{The concurrence $C_{3,4}^{1,2}$ as a function of time $t$ and the AC phase factor $\theta$ for $\beta=\arccos\frac{1}{2}$.}\label{Figure6}
\end{figure}

Now we compare the pairwise entanglement between chain with and without AC phase. To show the numerical result, we take the chain length N=6 and $\beta=\arccos(\frac{1}{2})$. Suppose at time $t=0$ the 1-th and 2-th spins are maximally entangled, according to Eq.(\ref{TransCon}), we get the concurrence dependence on phase $\theta$ and time $t$ between 3-th and 4-th site by numerical calculation, see Fig.\ref{Figure6}. The phase shift $\theta$ plays a positive role in entanglement transfer in our chain model that includes next nearest neighbouring interaction. The maximum concurrence $C_{3,4}^{1,2}$ for $\theta=0$ is $0.455$ at $t=14.704$, but when $\theta\neq0$ the maximum concurrence $C_{3,4}^{1,2}$ is not less than $0.982$(at $t=19.248$ and $\theta=2.130$), here we choose $t\in[0,20]$ and $\theta\in[-\pi,\pi]$.

\medskip

Now let us detect how the parameter $\beta$ influences the entanglement transfer. When A-C phase $\theta=0$, we suppose the initial entangled sites are $m_{1}=1$ and $m_{2}=2$ at time $t=0$ and set the total number of sites $N=4$, the coefficients are given by
\begin{eqnarray*}
&&|\alpha_{1}^{1,2}(t)|=\tfrac{1}{\sqrt{2}}|\cos[(\sqrt{2}\sin2\beta+2\cos^2\beta)t]|;\\
&&|\alpha_{2}^{1,2}(t)|=\tfrac{1}{\sqrt{2}}|\cos[(\sqrt{2}\sin2\beta+2\cos^2\beta)t]|;\\
&&|\alpha_{3}^{1,2}(t)|=\tfrac{1}{\sqrt{2}}|\sin[(\sqrt{2}\sin2\beta+2\cos^2\beta)t]|;\\
&&|\alpha_{4}^{1,2}(t)|=\tfrac{1}{\sqrt{2}}|\sin[(\sqrt{2}\sin2\beta+2\cos^2\beta)t]|
\end{eqnarray*}
The concurrence should be a periodic function for time $t$ when parameter $\beta$ is fixed.

\medskip

The concurrence $C_{3,4}^{1,2}$ oscillates with time for fixed parameter $\beta$. The oscillating frequency of the concurrence reaches the maximum for $\cos2\beta=\frac{\sqrt{3}}{3}$ and $\sin2\beta=\frac{\sqrt{6}}{3}$.

\medskip

Another interesting phenomena occurs for
\begin{equation}
\sqrt{2}\sin2\beta+2\cos^2\beta=0
\end{equation}
Then $\beta=\arccos(-\frac{\sqrt{6}}{3})$, $|\alpha_{1}^{1,2}(t)|=|\alpha_{2}^{1,2}(t)|=\frac{1}{\sqrt{2}}$, $|\alpha_{3}^{1,2}(t)|=|\alpha_{4}^{1,2}(t)|=0$, the coefficients make the entanglement transfer unable in our chain model for $N=4$ due to the effect of the interactions of $n,n+1$-site and $n,n+2$-th site. $\beta=\arccos(-\frac{\sqrt{6}}{3})$  is exactly a sufficient condition for generating genuine entangled 3-qubit pure states shown in Sec.\ref{Sec:2}.

\section{Conclusion and Discussion}
\label{Sec:Conclusion}

In this paper, we first show that 3-body scattering matrix of Yang-Baxter equation really generates entangled 3-qubit state by acting $\breve{R}_{123}(\eta,\beta,\varphi)$ on direct product state and investigate the relation between the parameters $\eta$, $\beta$ in 3-body S-Matrix and the corresponding concurrences. We give the condition for generating two type of genuine entangled pure states from direct states by 3-body S-Matrix transformation: GHZ states are generated when $|\cos\eta|=\frac{1}{2}$ and $|\cos\beta|=\frac{\sqrt{6}}{3}$, W states are generated when $|\cos\eta|=0$ and $|\cos\beta|=\frac{\sqrt{6}}{3}$. We also find that the bipartite concurrence of 3-qubit pure state can be expressed by 2-qubit concurrence(relate to 2-body S-matrix).  Second, we obtain the Hamiltonian based on the S-Matrix and the Berry phase in 3-qubit entanglement space that depends on one parameter $\eta$ only. The solid angle enclosed by the loop on the Bloch sphere is realized in terms of $\eta$. Third, we construct a 3-body S-matrix Hamiltonian chain model and find it is the extension of 1D Kitaev toy model, and show that the YBE condition kills the unpaired Majorana fermions at the end of the chain. We show that the parameters in entanglement space $\eta=-\frac{\pi}{3}$ and $\beta=\arccos\frac{\sqrt{6}}{3}$ corresponds to the ``Phase Boundary'' $2\omega=\mu$ proposed by Kitaev, meanwhile the parameter $\eta=-\frac{\pi}{3}$ and $\beta=\arccos\frac{\sqrt{6}}{3}$ is also a sufficient condition for generating 3-qubit genuine entangled GHZ states in Sec.\ref{Sec:2}. Forth, we consider the special case of the chain model in $\eta=\frac{\pi}{2}$. The boost of AC effect in entanglement transfer is discussed, and we find that when $\beta=\arccos(-\tfrac{\sqrt{6}}{3})$ for the chain length N=4, the entanglement cannot be transferred with time $t$ although there are interactions between different spin sites due to YBE condition.
So we guess that there is much closer relation between 3-body S-matrix factorized by Yang-Baxter Equation and 3-qubit pure state entanglement as well as the chain model for 3-body S-matrix to be detected.
\section*{Acknowledgments}
We thank R. Y. Zhang for helpful discussions. This work is in part supported by NSF of China (Grants No. 11275024 and No. 11075077)
\appendix

\section*{Appendix A: Calculation of 3-body S-matrix in the exponential form }
3-body S-matrix is expressed as:
\begin{eqnarray}
 \breve{R}_{123}(\theta_1,\theta_2,\theta_3,\varphi)&=&\breve{R}_{12}(\theta_1,\varphi) \breve{R}_{23}(\theta_2,\varphi)\breve{R}_{12}(\theta_3,\varphi)\nonumber\\
 &=&[\breve{R}\otimes I][I\otimes \breve{R}][\breve{R}\otimes I]\label{App3Scat}
\end{eqnarray}
where
\begin{equation}
\breve{R}=\cos\theta_{i}I\otimes I+i\sin\theta_{i}\sigma_2\otimes\sigma_1
\end{equation}
with $\sigma_1=\left(\begin{array}{cc}
  0 & e^{i\varphi} \\
  e^{-i\varphi} & 0
  \end{array}\right)
$, $\sigma_2=\left(\begin{array}{cc}
  0 & -ie^{i\varphi} \\
  ie^{-i\varphi} & 0
  \end{array}\right)$, $\sigma_3=\left(\begin{array}{cc}
  1 & 0 \\
  0 & -1
  \end{array}\right)$, they satisfy $[\sigma_i, \sigma_j]=i\epsilon_{ijk}\sigma_k$.

Then Eq.(\ref{App3Scat}) is expressed by:(denote $a\otimes b\otimes c\equiv abc$)
\begin{eqnarray*}
& &\breve{R}_{123}(\theta_1,\theta_2,\theta_3,\varphi)\\
&=&[\cos\theta_1 I^{\otimes3}+i\sin\theta_1\sigma_2\sigma_1 I][\cos\theta_2 I^{\otimes3}+i\sin\theta_2 I \sigma_2\sigma_1]\\
& &[\cos\theta_3 I^{\otimes3}+i\sin\theta_3\sigma_2\sigma_1I]\\
&=&\cos\theta_2\cos(\theta_1+\theta_3)I^{\otimes3}-i\sin\theta_2\sin(\theta_1-\theta_3)\sigma_2\sigma_3\sigma_1\\
& &+i\cos\theta_2\sin(\theta_1+\theta_3)\sigma_2\sigma_1I+i\sin\theta_2\cos(\theta_1-\theta_3)I\sigma_2\sigma_1
\end{eqnarray*}
With the deformation of the relation(Eq.\ref{Anglerelation}) between $\theta_1$, $\theta_2$ and $\theta_3$:
\begin{equation}
\sin\theta_2\cos(\theta_1-\theta_3)=\cos\theta_2\sin(\theta_1+\theta_3)
\end{equation}
We have
\begin{eqnarray*}
& &\breve{R}_{123}(\theta_1,\theta_2,\theta_3,\varphi)\\
&=&[\cos\theta_1 I^{\otimes3}+i\sin\theta_1I\sigma_2\sigma_1][\cos\theta_2 I^{\otimes3}+i\sin\theta_2\sigma_2\sigma_1 I]\\
& &[\cos\theta_3 I^{\otimes3}+i\sin\theta_3I\sigma_2\sigma_1]\\
&=&\cos\theta_2\cos(\theta_1+\theta_3)I^{\otimes3}-i\sin\theta_2\sin(\theta_1-\theta_3)\sigma_2\sigma_3\sigma_1\\
& &+i\sin\theta_2\cos(\theta_1-\theta_3)\sigma_2\sigma_1I+i\sin\theta_2\cos(\theta_1-\theta_3)I\sigma_2\sigma_1\\
&=&\cos\eta I^{\otimes 3}+i\sin\eta(\vec{n}\cdot \vec{\Sigma})
\end{eqnarray*}
where:
\begin{eqnarray}
&&\cos\eta=\cos\theta_2\cos(\theta_1+\theta_3)\\
&&\sin\eta=\sin\theta_2\sqrt{1+\cos^2(\theta_1-\theta_3)}\\
&&\cos\beta=\tfrac{\sqrt{2}\cos(\theta_1-\theta_3)}{\sqrt{1+\cos^2(\theta_1-\theta_3)}}\\
&&\sin\beta=\tfrac{\sin(\theta_3-\theta_1)}{\sqrt{1+\cos^2(\theta_1-\theta_3)}}\\
&&\vec{n}=(\tfrac{1}{\sqrt{2}}\cos\beta, \tfrac{1}{\sqrt{2}}\cos\beta, \sin\beta)\\
&&\vec{\Sigma}=(\sigma_2\sigma_1 I, I\sigma_2\sigma_1, \sigma_2\sigma_3\sigma_1)
\end{eqnarray}
It is easy to check that $(\vec{n}\cdot \vec{\Sigma})^2=I^{\otimes 3}$, then we have
\begin{equation}
\breve{R}_{123}(\theta_1,\theta_2,\theta_3,\varphi)=\cos\eta I^{\otimes 3}+i\sin\eta(\vec{n}\cdot \vec{\Sigma})=e^{i\eta \left(\vec{n}\cdot\vec{\Sigma}\right)}
\end{equation}

\section*{Appendix B: CALCULATIONS OF ``ZERO MODE'' IN SECTION \ref{sec:4} }
From Eq.(\ref{3Bparameter}), with the assumption $\omega_2=\Delta_2$  and the parameters' relation(Eq.(\ref{3Bparameter})), all of the six parameters have a common factor $\omega_2$:
\begin{subequations}
\begin{align*}
\omega_1&=\sqrt{2}\tan\beta \omega_2\\
\Delta_1&=-\frac{1}{\sqrt{2}\tan\beta} \Delta_2=-\frac{1}{\sqrt{2}\tan\beta} \omega_2\\
\mu_1&=-(1+2\tan^2\beta)\omega_2\\
\mu_2&=-2\omega_2\\
\omega_2&=\Delta_2
\end{align*}
\end{subequations}
 Let $\tan\beta\equiv b$, we find that $\hat{H_3}$ is given by
\begin{equation}
\hat{H}_3=\frac{i\omega_2}{4}\sum_{i,j}A_{ij}c_{i}c_{j}
\end{equation}
with the matrix
\begin{widetext}
\begin{subequations}
\begin{align*}
A=\left[
   \begin{array}{cccccccccc}
   0 & 1+2b^2 &  0 & \frac{-2b^2-1}{\sqrt{2}b} &  0 & 0 &  0 &  0 & \cdots&  0  \\
   -1-2b^2 &  0 & \frac{2b^2-1}{\sqrt{2}b} &  0 &  2 &  0 &  0 &  0 &  \cdots &  0\\
   0 & \frac{-2b^2+1}{\sqrt{2}b} &  0 &  3+2b^2 &  0 & \frac{-4b^2-2}{\sqrt{2}b} &  0 &  0 & \cdots & 0 \\
     \frac{2b^2+1}{\sqrt{2}b} &  0 & -3-2b^2 &  0 &  \frac{4b^2-2}{\sqrt{2}b} & \ 0 &  2 & 0 & \cdots & 0\\
     \\
     0 & -2 & 0 & \frac{-4b^2+2}{\sqrt{2}b} &  0 &  4b^2+4 &  0 & \frac{-4b^2-2}{\sqrt{2}b} &  0 & \cdots \\
     0 &  0 & \frac{4b^2+2}{\sqrt{2}b} &  0 &  -4b^2-4 & \ 0 &  \frac{4b^2-2}{\sqrt{2}b} & 0 & 2 & \cdots\\
     \vdots & \vdots & \vdots & \vdots & \vdots & \vdots & \vdots & \vdots \\
     \vdots & \vdots & \vdots & \vdots & \vdots & \vdots & \vdots & \vdots \\
     \vdots & \vdots & \vdots & \vdots & \vdots & \vdots & \vdots & \vdots \\
     \cdots &  -2 & 0 &  \frac{-4b^2+2}{\sqrt{2}b} &  0 &  4b^2+4 & 0 & \frac{-4b^2-2}{\sqrt{2}b} & 0 & 0\\
     \cdots & 0 & \frac{4b^2+2}{\sqrt{2}b} & 0 &  -4b^2-4 & 0 &  \frac{4b^2-2}{\sqrt{2}b} & 0 &  2 & 0 \\
     \\
     0 &  \cdots & 0 &  -2 &  0 & \frac{-4b^2+2}{\sqrt{2}b} &  0 & 3+2b^2 & 0 & \frac{-2b^2-1}{\sqrt{2}b}\\
     0 & \cdots & 0 &  0 &  \frac{4b^2+2}{\sqrt{2}b} &  0 & -3-2b^2 &  0 & \frac{2b^2-1}{\sqrt{2}b} & 0 \\
     0 & \cdots & 0 & 0 & 0 & -2 &  0 & \frac{-2b^2+1}{\sqrt{2}b} & 0 & 1+2b^2\\
     0 & \cdots & 0 & 0 & 0 &  0 & \frac{2b^2+1}{\sqrt{2}b} & 0 & -1-2b^2 & 0
   \end{array} \right]
\end{align*}
\end{subequations}
\end{widetext}
 In matrix A,  there are only 4 non-zero elements in each row(from 5-th to (2N-4)-th row) and the other entries can be viewed as boundary condition. If there exist zero eigenvalues for matrix A, the eigenvector should be
 \begin{subequations}
 \begin{align}
 &|V_1\rangle=\alpha'_1 |x_1\rangle+\alpha'_2 |x_2\rangle+\alpha'_3 |x_3\rangle\\
 &|V_2\rangle=\alpha''_1 |x^{-1}_1\rangle+\alpha''_2 |x^{-1}_2\rangle+\alpha''_3 |x^{-1}_3\rangle
\end{align}
\end{subequations}
with ($i=1,2,3$)
 \begin{subequations}\label{solutionofA}
 \begin{align}
&|x_i\rangle=(0, x_i, 0, x_i^2, 0, \cdots 0, x_i^{N-1}, 0, x_i^{N} )\\
&|x_i^{-1}\rangle=( x_i^{-1}, 0, x_i^{-2}, 0, \cdots  x_i^{-N+1}, 0, x_i^{-N}, 0)
\end{align}
\end{subequations}
Substitute them into matrix A, $x_i$ are the roots of
 \begin{eqnarray}
&&(2\tan^2\beta+1)x^3-2\sqrt{2}(\tan^3\beta+\tan\beta)x^2\nonumber\\
&&\quad+(2\tan^2\beta-1)x+\sqrt{2}\tan\beta=0
\end{eqnarray}
 $\alpha_{i}$ and $\alpha'_{i}$ are used for satisfying the boundary condition in 1 to 4-th row and (2N-3) to 2N-th row of matrix A.

\bibliographystyle{unsrt}
\bibliography{shanjianDoc9}
\end{document}